\documentclass{article}

\usepackage{jcappub}
\usepackage[utf8]{inputenc}
\usepackage{amsmath,amsfonts,amssymb}
\usepackage{graphicx}
\usepackage{subcaption}
\usepackage[toc,page]{appendix}
\usepackage{cleveref}
\usepackage[loose]{units}

\newcommand{\beqn}{\begin{equation}}
\newcommand{\eeqn}{\end{equation}}
\newcommand{\dd}{\mathrm{d}}
\newcommand{\gmn}{g_{\mu\nu}}
\newcommand{\fmn}{f_{\mu\nu}}

\newcommand{\mFP}{m_\mathrm{FP}}
\newcommand{\OFP}{\Omega_\mathrm{FP}}
\newcommand{\OL}{\Omega_\Lambda}
\newcommand{\Om}{\Omega_{\mathrm{m,0}}}
\newcommand{\OFPN}{\Omega_\mathrm{FP,0}}
\newcommand{\OLN}{\Omega_{\Lambda,0}}

\begin{document}

\begin{flushright} 
\texttt{ MPP-2020-27 }
\end{flushright}

\title{Physical parameter space of bimetric theory and SN1a constraints}

\author[a]{Marvin L{\"u}ben,}

\author[a]{Angnis Schmidt-May,}

\author[b,c,d]{Jochen Weller}

\affiliation[a]{Max-Planck-Institut f\"ur Physik (Werner-Heisenberg-Institut),\\
 F\"ohringer Ring 6, 80805 Munich, Germany}
\affiliation[b]{Universit\"ats-Sternwarte, Ludwig-Maximilians-Universit\"at M\"unchen\\
Scheinerstr. 1, 81679 Munich, Germany}
\affiliation[c]{Excellence Cluster Origins, Bolzmannstr. 2, D-85748 Garching, Germany}
\affiliation[d]{Max Planck Institute for Extraterrestrial Physics, \\
Giessenbachstr. 1, 85748 Garching, Germany}

\emailAdd{mlueben@mpp.mpg.de}
\emailAdd{angnissm@mpp.mpg.de}
\emailAdd{jochen.weller@usm.lmu.de}

\abstract{
Bimetric theory describes a massless and a massive spin-2 field with fully
non-linear (self-)interactions. It has a rich phenomenology and has been
successfully tested with several data sets. However, the observational 
constraints have not been combined in a consistent framework, yet.
We propose a parametrization of bimetric solutions in terms of the effective
cosmological constant $\Lambda$ and the mass $m_{\rm FP}$ of the spin-2 field as well as
its coupling strength to ordinary matter $\bar\alpha$.
This simplifies choosing priors in statistical analysis and
allows to directly constrain these parameters with observational data
not only from local systems but also from cosmology.
By identifying the physical vacuum of bimetric theory these parameters are uniquely
determined.
We work out the new parametrization for various submodels and
present the implied consistency constraints on the physical parameter space.
As an application we derive observational constraints from
SN1a on the physical parameters.
We find that a large portion of the physical parameter space is
in perfect agreement with current supernova data
including self-accelerating models with a heavy spin-$2$ field.
}


\setcounter{tocdepth}{2}
\maketitle

\section{Introduction}

The Standard Model of particle physics contains particles with different spin
numbers up to $1$.
For each spin number there are consistent (field) theories describing
massless and massive particles.
Going higher in the spin number, the theory of General Relativity
contains a spin-$2$ field that is massless, the (yet unobserved) graviton.
The question arises whether one can construct
a consistent theory describing a spin-$2$ field that is massive.
The first attempt was undertaken by Fierz and Pauli in
1939 who proposed a linear
theory~\cite{Pauli:1939xp,Fierz:1939ix}.
Boulware and Deser argued that any non-linear completion
of the linearized theory must contain a
ghost~\cite{vanDam:1970vg,Zakharov:1970cc,Vainshtein:1972sx,Boulware:1973my}.
However, in 2010/11 a ghost-free and fully non-linear theory
describing a massive spin-$2$ field in flat spacetime was
presented usually referred to as
(dRGT) massive
gravity~\cite{deRham:2010ik,deRham:2010kj,Hassan:2011vm,Hassan:2011hr}.
Hassan and Rosen generalized the theory to bimetric
theory
that describes a gravitating massive spin-$2$
field~\cite{Hassan:2011zd,Hassan:2011ea}.
Massive gravity and bimetric theory hence fill the gap in the list
of consistent field theories describing
massless and massive particles with spin up to $2$.
For a review on bimetric theory we refer to Ref.~\cite{Schmidt-May:2015vnx}.

The massive spin-$2$ field has various phenomenological implications,
from local to cosmological scales.
Bimetric theory has cosmological solutions which give rise to an accelerated expansion
of the universe at late times even in the absence of vacuum energy~\cite{vonStrauss:2011mq,Konnig:2013gxa,Akrami:2012vf,DeFelice:2014nja}. This feature is
usually referred to as self-acceleration. Within bimetric theory the interaction energy
between the spin-$2$ fields is responsible for the late time acceleration, besides a
possible vacuum energy component.
Moreover, bimetric theory contains a Dark Matter candidate, the massive
spin-$2$ field~\cite{Babichev:2016hir,Babichev:2016bxi,Chu:2017msm}.
On galaxy cluster to galactic scales, the fifth force mediated by the massive spin-$2$ field
gives rise to beneficial deviations from General Relativity,
affecting the required Dark Matter abundance in
these systems~\cite{Platscher:2018voh}.
On smaller scales the fifth force is suppressed due the Vainshtein screening
mechanism~\cite{Vainshtein:1972sx,Babichev:2013pfa} as demanded by local
tests of gravity~\cite{Will:2014kxa}.
This, however, depends on the mass of the spin-$2$ field and its coupling strength
to ordinary matter.
A collection of various phenomenological features of bimetric theory
can be found, e.g., in Ref.~\cite{Luben:2018ekw}.

In this paper we aim at constraining the physical parameter space of bimetric theory
with cosmological data. By physical parameters
we mean, e.g., the mass of the massive spin-$2$ field, $\mFP$, and its coupling
strength to matter, $\bar\alpha$.
Of course, bimetric theory has been compared to cosmological data previously
on both the background and perturbative level, see ,e.g.,
Refs.~\cite{vonStrauss:2011mq,Konnig:2013gxa,Akrami:2012vf,DeFelice:2014nja,Solomon:2014dua,Konnig:2014xva,Mortsell:2018mfj,Lindner:2020eez}.
All these studies found various (sub-)models that give rise to a viable background cosmology
which can compete with General Relativity.
However, the existing results cannot be combined in a straightforward manner due to
different parameterizations and assumptions.

To translate the observational constraints coming from cosmology to the
physical parameters, these must be related to the parameters of the theory in a unique way.
This requires to identify the physical vacuum out of the up to four vacuum solutions
of bimetric theory. We do this by imposing theoretical consistency requirements on the
vacuum and on the cosmic expansion history. Further we demand that the consistent vacuum
corresponds to the infinite future of the viable expansion history which identifies the
physical vacuum of bimetric theory.
The first aim of this paper is to identify the physical vacuum for all bimetric models
with up to three non-vanishing interaction parameters and work out the dictionary
between the different parameterizations. Thereby we find theoretical consistency
constraints on the physical parameter space.
The second aim is to apply this procedure
to constrain the physical parameters of these (sub-)models with
data from supernovae of Type Ia.

The paper is organized as follows. In \cref{sec:introduction} we give a brief introduction
to bimetric theory and in \cref{sec:bkg-cosmo} we review FLRW solutions.
In~\cref{sec:vacuum-solutions}
we propose the physical parametrization and explain how to construct it.
While in \cref{sec:submodels} we apply the procedure to various (sub-)models of bimetric theory,
we perform the data analysis for each (sub-)model in \cref{sec:obs-constraints}.
Finally, in \cref{sec:outlook} we summarize our results and discuss possible next steps.

\section{Bimetric theory in a brief}\label{sec:introduction}

In this section, we summarize those aspects of bimetric theory needed
to study cosmological solutions.
After presenting the action and the equations of motion, we discuss 
vacuum solutions and the mass spectrum of the linearized theory.

\subsection{Action and equations of motion}

We focus on a version of bimetric theory (so-called singly-coupled) where matter fields 
couple minimally to only one of the metric tensors\footnote{
To which metric the matter fields should couple led to a lot of discussion in the
literature \cite{deRham:2014naa,Yamashita:2014fga,Luben:2018kll,deRham:2014fha,Akrami:2014lja,Hinterbichler:2015yaa}.
The result is that there are only two options which do not reintroduce the Boulwere-Deser
ghost at unacceptable low energy scales.
A matter field can minimally couple to only one of the metric tensors. That allows for
two independent matter sectors (one for $\gmn$ and one for $\fmn$) that do not couple 
directly to each other, but only via their gravitational interactions 
\cite{Yamashita:2014fga,deRham:2014naa}. 
The singly-coupled version in which we work in this article is a special case without 
a $\fmn$-matter sector.
Alternatively, matter can minimally couple to an effective metric composed out
of the two metric tensors \cite{deRham:2014naa}.
This matter coupling lowers the cutoff of the theory
but can be embedded into a trimetric setup~\cite{Luben:2018kll}.
Phenomenological aspects were discussed in e.g. Refs.~\cite{Enander:2014xga,Solomon:2014iwa,Gumrukcuoglu:2015nua,Gumrukcuoglu:2014xba,Aoki:2013joa,DeFelice:2014nja}.}
, say $\gmn$.
The ghost-free action is given by
\cite{deRham:2010ik,deRham:2010kj,Hassan:2011hr,Hassan:2011vm,Hassan:2011tf,Hassan:2011zd}
\begin{flalign}
	S=&m_g^2 \int\dd^4 x\left[ \sqrt{-g}\, R(g)+\alpha^2\sqrt{-f}\, R(f)-2\sqrt{-g}\, V(g,f)\right]\nonumber\\
	&+\int\dd^4x\sqrt{-g}\, \mathcal L_\mathrm{m}(g,\phi_i)\label{eq:bimetric-action}
\end{flalign}
where $R(g)$ and $R(f)$ are the Ricci scalars of the two 
metric tensors $g_{\mu\nu}$ and $f_{\mu\nu}$, resp.
The parameter $m_g$ is the Planck mass of $g_{\mu\nu}$ 
and the quantity $\alpha$ measures the ratio of $m_g$ to the 
$\fmn$-Planck mass.
The metric tensors interact via the bimetric potential
\begin{flalign}\label{eq:pot-interchange-symmetry}
	V( g , f ; \beta_n )=\sum_{n=0}^4\beta_n e_n(S)\,,
\end{flalign}
which is defined in terms of the elementary symmetric polynomials $e_n$ \cite{Hassan:2011vm}.
These are functions of the square-root matrix $S$ defined as
\begin{flalign}
	S^\mu_{\ \alpha}S^{\alpha}_{\ \nu}=g^{\mu\alpha}f_{\alpha\nu}
\end{flalign}
or in matrix notation, $S=\sqrt{g^{-1}f}$.
Due to the properties of the elementary symmetric polynomials, the bimetric potential
satisfies the relation
\begin{flalign}
	\sqrt{-g}\, V(g,f;\beta_n)=\sqrt{-f}\, V(f,g;\beta_{4-n}).
\end{flalign}
The interaction parameters $\beta_n$ are 
constant parameters of mass dimension $2$ (in our normalization),
where $\beta_0$ parametrizes the vacuum energy for 
$\gmn$ and similarly $\beta_4$ for $\fmn$.
Matter fields, which we collectively denote as $\phi_i$ couple minimally 
to $\gmn$ and $\mathcal L_\mathrm{m}$ is some generic
matter Lagrangian.
Therefore, $\gmn$ is the physical metric which defines the geometry
in which the matter fields $\phi_i$ live.

When varying the action~\eqref{eq:bimetric-action} w.r.t $g^{\mu\nu}$ and $f^{\mu\nu}$,
we arrive at two sets of Einstein field equations:
\begin{flalign}
	G_{\mu\nu}+V_{\mu\nu}=\frac{1}{m_g^2} T_{\mu\nu} \ , \ \ \
	\tilde G_{\mu\nu}+\frac{1}{\alpha^2} \tilde V_{\mu\nu}=0\label{eq:einstein-eq}
\end{flalign}
where $G_{\mu\nu}$ and $\tilde G_{\mu\nu}$ are the usual Einstein tensors of $\gmn$ and $\fmn$, resp.
The stress-energy tensor of matter is given by
\begin{flalign}
	T_{\mu\nu}=-\frac{2}{\sqrt{-g}}\frac{\delta \sqrt{-g}\mathcal L_{\rm m}}{\delta g^{\mu\nu}}.
\end{flalign}
The contributions from the bimetric potential are given by
\begin{flalign}
	V_{\mu\nu}=\sum_{n=0}^3(-1)^n \beta_n g_{\mu\lambda} Y^\lambda_{(n)\nu}(S) \ , \ \ \ 
	\tilde V_{\mu\nu}=\sum_{n=0}^3(-1)^n \beta_{4-n} f_{\mu\lambda} Y^\lambda_{(n)\nu}(S^{-1})
\end{flalign}
where the explicit form of the function $Y^\lambda_{(n)\nu}$ can be found in, e.g.,
Ref.~\cite{Hassan:2011vm}.

Both Einstein tensors satisfy the Bianchi identities, $\nabla^\mu G_{\mu\nu}=0$ and 
$\tilde\nabla^\mu \tilde G_{\mu\nu}=0$, where 
$\nabla^\mu$ is 
the covariant derivative compatible with $g_{\mu\nu}$ and $\tilde \nabla^\mu$ is the 
covariant derivative compatible with $f_{\mu\nu}$.
If the matter action is invariant under diffeomorphisms, its stress-energy tensor 
satisfies the conservation equation 
\begin{flalign}\label{eq:SE-conservation}
	\nabla^\mu T_{\mu\nu}=0\,.
\end{flalign}
This results in the \textit{Bianchi constraint} in bimetric theory,
\begin{flalign}\label{eq:bianchi-constraint}
	\nabla^\mu  V_{\mu\nu}=0\ , \ \ \
	\tilde\nabla^\mu \tilde V_{\mu\nu}=0 \,,
\end{flalign}
where it can be shown that one equation implies the other due to diffeomorphism
invariance.

At this stage, we already note that the action~\eqref{eq:bimetric-action} is invariant 
under the map,
\begin{flalign}\label{eq:Z2-symmetry}
	\sqrt{g^{-1}f} \ \longrightarrow \ -\sqrt{g^{-1}f} \ \ , \ \ \beta_n \ \longrightarrow \ (-1)^n\beta_n \,.
\end{flalign}
Suppose, $S=\sqrt{g^{-1}f}$ is a solution to the bimetric field 
equations with interaction parameters $\beta_n$.
This solution is dual to the solution $-S$ with interaction parameters $(-1)^n\beta_n$.
Hence, considering only half of the solutions
already covers the entire solution space for arbitrary interaction parameters.
We will come back to this point when studying cosmological solutions.
Note that for $\beta_1=\beta_3=0$,~\cref{eq:Z2-symmetry} is a symmtery of the theory.

\subsection{Proportional backgrounds and mass spectrum}

After having presented the bimetric field equations, let us study an important
class of solutions: the proportional background.
Let both metrics be related by a conformal factor $c$ as
\begin{flalign}
	\bar f_{\mu\nu}=c^2 \bar g_{\mu\nu} \,.
\end{flalign}
The Bianchi constraint forces $c$ to be a constant.
The field equations reduce to two sets of Einstein equations,
\begin{flalign}
	G_{\mu\nu}(\bar g_{\mu\nu}) + \Lambda_g \bar g_{\mu\nu} =0 \ , \ \ \ 
	\tilde G_{\mu\nu}(\bar f_{\mu\nu}) + c^{-2} \Lambda_f \bar f_{\mu\nu} =0\,.
\end{flalign}
The cosmological constants $\Lambda_g$ and $\Lambda_f$ originate from the 
bimetric potential and are given by
\begin{subequations}\label{eq:eff-cosmological-constant}
\begin{flalign}\label{eq:eff-cosmological-constant-g}
	&\Lambda_g =\beta_0 + 3\beta_{1} c + 3\beta_{2} c^2 + \beta_3 c^3\\
	&\Lambda_f =\frac{1}{\alpha^2c^2}\left(\beta_1 c + 3\beta_2 c^2 + 3\beta_3 c^3 + \beta_4 c^4\right).
\end{flalign}
\end{subequations}
Since $G_{\mu\nu}(\bar g_{\mu\nu})=\tilde G_{\mu\nu}(\bar f_{\mu\nu})$,
combining the Einstein equations results in $\Lambda_g=\Lambda_f\equiv\Lambda$,
which explicitly reads,
\begin{flalign} \label{eq:cquartic}
	\alpha^{2}\beta_{3}c^{4}+(3\alpha^{2}\beta_{2}-\beta_{4})c^{3}+3(\alpha^{2}\beta_{1}-\beta_{3})c^{2} +(\alpha^{2}\beta_{0}-3\beta_{2})c-\beta_{1}=0.
\end{flalign}
This is a polynomial in $c$ and has up to four real-valued roots, which determine
$c$ in terms of the bimetric parameters, $c=c(\alpha,\beta_{n})$.
The proportional backgrounds exist only in vacuum; 
matter stress-energy drives the solution away from the
proportional background.
Hence, each root $c$ corresponds to a vacuum of bimetric theory.
We will study these vacua in more detail in~\cref{sec:vacuum-solutions}.

Depending on the sign of the effective cosmological constant $\Lambda$,
the proportional background can describe (Anti-)de Sitter or Minkowski space.
Only in such spacetimes a well-defined notion of spin
and mass exists due to the presence of Poincare invariance.
To find the mass spectrum, we study linear fluctuations around the proportional background,
\begin{flalign}\label{eq:proportional-background}
	g_{\mu\nu}=\bar g_{\mu\nu}+\frac{1}{m_g}\delta g_{\mu\nu}\ , \ \ \
	f_{\mu\nu}=c^2 \bar g_{\mu\nu}+\frac{c}{m_f}\delta f_{\mu\nu}\,,
\end{flalign}
where $\delta g_{\mu\nu}$ and $\delta\fmn$ are the canonically normalized
linear fluctuations.
The mass eigenstates are given by a linear combination of the metric
fluctuations~\cite{Hassan:2012wr}
\begin{flalign}
	\delta G_{\mu\nu}=\frac{1}{\sqrt{1+\alpha^2c^2}}(\delta g_{\mu\nu}+\alpha c\,\delta f_{\mu\nu}) \ , \ \ \
	\delta M_{\mu\nu}=\frac{1}{\sqrt{1+\alpha^2c^2}}(\delta f_{\mu\nu}-\alpha c\,\delta g_{\mu\nu})\,,
\end{flalign}
where the mode $\delta G_{\mu\nu}$ describes a massless spin $2$-field
and the mode $\delta M_{\mu\nu}$ a massive spin $2$-field.
Its mass in terms of the bimetric parameters is given by
\begin{flalign}\label{eq:fierz-pauli-mass}
	\mFP^2=\left(1+\frac{1}{\alpha^2 c^2}\right)(\beta_1c+2\beta_2 c^2+\beta_3 c^3)\,.
\end{flalign}
The metric fluctuations are a linear superposition of the mass eigenstates,
\begin{flalign}
	\delta g_{\mu\nu}= \frac{1}{\sqrt{1+\alpha^2c^2}} (\delta G_{\mu\nu}-\alpha c\, \delta M_{\mu\nu})\ , \ \ \
	\delta f_{\mu\nu}= \frac{1}{\sqrt{1+\alpha^2c^2}}(\delta M_{\mu\nu}+\alpha c\, \delta G_{\mu\nu}).
\end{flalign}
This allows for a physical interpretation of the combination $\alpha c$.
It measures the mixing of the mass eigenstates in the original
metric fluctuations and can be thought of as being related to a mixing angle.
In the limit $\alpha c\rightarrow 0$, the massive mode drops out of the fluctuations of
the physical metric $\gmn$.
Since in singly-coupled bimetric theory matter couples to $\gmn$,
we expect to recover the laws of GR in that limit~\cite{Akrami:2015qga}.

In de Sitter space, unitarity forbids the mass of the spin-$2$ field to be arbitrarily small.
The mass has to satisfy the Higuchi bound \cite{Higuchi:1986py,Higuchi:1989gz}
\begin{flalign}\label{eq:Higuchi-bound}
	m_\mathrm{FP}^2 \geq \frac{2}{3}\Lambda\,,
\end{flalign}
in order to ensure that the helicity-$0$ mode of the massive spin $2$-field
is not a ghost state (Higuchi ghost) in the sense that its kinetic term
has the correct sign.

\section{Background cosmology}\label{sec:bkg-cosmo}

\subsection{Flat FLRW Ansatz}

After having introduced bimetric theory and discussed vacuum solutions
we now consider cosmological solutions.
Following the cosmological principle, we assume spacetime to be 
homogenous and isotropic on large scales and spatially flat.
Both metrics assume the flat FLRW form,
\begin{flalign}
	\dd s_g^2 & = -\dd t^2+a(t)^2\dd x^2\,,\\
	\dd s_f^2 & = -X(t)^2\dd t^2+b(t)^2\dd x^2\,,
\end{flalign}
where $a$ and $b$ are the scale factors of the metric $\gmn$ and $\fmn$ resp., and $X$
is the lapse of the metric $\fmn$. We used the time 
reparametrization-invariance already to set the lapse of the metric $\gmn$ to unity. This 
fixes the gauge completely.
From now on we do not explicitly write the time-dependence of the metric functions.
For later let us define the ratio 
of the scale factors and the Hubble rates as
\begin{flalign}
	y=\frac{b}{a} \,, \ \ H=\frac{\dot a}{a} \,, \ \ H_f=\frac{\dot b}{Xb} \,,
\end{flalign}
where a dot represents derivative w.r.t. cosmic time $t$.
According to homogeneity and isotropy, we assume the universe to be filled with a 
perfect fluid with stress-energy tensor
\begin{flalign}
	T_{\mu\nu}=(\rho_\mathrm{m}+p_\mathrm{m})u_\mu u_\nu + p_\mathrm{m}\, g_{\mu\nu}\,,
\end{flalign}
where $u_\mu$ is the $4$-velocity of the fluid with energy density $\rho_\mathrm{m}$
and pressure $p_\mathrm{m}$. The latter quantities are related via the linear equation
of state
\begin{flalign}
	w_\mathrm{m}=\frac{p_\mathrm{m}}{\rho_\mathrm{m}}\,.
\end{flalign}
In this work we are mostly interested in the late-time behavior of the universe and
in particular in times after radiation-matter-equality. Later we 
will set $w_\mathrm{m}=0$ in order to describe non-relativistic matter such as
baryons and dark matter.

\subsection{Equations of motion}

The Bianchi constraint \eqref{eq:bianchi-constraint} on the FLRW ansatz reads
\begin{flalign}
	\left(\dot b-X \dot a\right)\left(\beta_1+2\beta_2 y+\beta_3 y^2\right)=0\,.
\end{flalign}
There are two branches of solutions.
Either one demands the term in the second parentheses to vanish,
which forces the ratio of the scale factors to be a constant, i.e. $y=\text{const}$.
This algebraic branch 
is pathological \cite{Comelli:2012db,Cusin:2015tmf} and implies that the mass of the
massive spin-$2$ field is identically zero, cf.~\cref{eq:fierz-pauli-mass}.
The other solution is given by demanding the term in the first parentheses
to vanish, which reads in terms of the Hubble rates
\begin{flalign}
	H = y H_f \,.
\end{flalign}
This solution is referred to as dynamical branch on which we will focus for the remainder
of the paper.

For the isotropic and homogenous ansatz, the conservation \cref{eq:SE-conservation}
reduces to the continuity equation
\begin{flalign}\label{eq:continuity}
	\dot\rho_\mathrm{m} + 3H(1+w_\mathrm{m})\rho_\mathrm{m}=0\,.
\end{flalign}
This equation is solved by
\begin{flalign}\label{eq:matter-conservation}
	\rho_\mathrm{m}=\rho_{\mathrm{m},0} a^{-3(1+w_{\rm m})}\,,
\end{flalign}
where $\rho_{\mathrm{m},0}$ is a constant of integration.
The time-time-component of the Einstein field \cref{eq:einstein-eq} for
$\gmn$ and $\fmn$ on the dynamical branch read
\begin{subequations}\label{eq:mod-friedmann}
\begin{flalign}
	3H^2 & = \frac{1}{m_g^2}\left(\rho_\mathrm{DE}+\rho_\mathrm{m}\right)\, \label{eq:mod-friedmann-g}\\
	3H^2 & = \frac{1}{m_g^2} \rho_\mathrm{pot}\,,\label{eq:mod-friedmann-f}
\end{flalign}
\end{subequations}
where we have defined the energy densities coming from the interaction potential as
\begin{subequations}
\begin{flalign}
	\rho_\mathrm{DE} &=m_g^2\left(\beta_0+3\beta_1 y + 3\beta_2 y^2 +\beta_3 y^3\right)\,,\\
	\rho_\mathrm{pot} &= \frac{m_g^2}{\alpha^2 y^2}\left(\beta_1 y +3\beta_2 y^2+ 3\beta_3 y^3 + \beta_4 y^4\right)\,.
\end{flalign}
\end{subequations}
Both energy densities are time dependent via $y$.
The effect of the interaction potential can be interpreted as dynamical
Dark Energy with a non-constant equation of state.
The \cref{eq:continuity,eq:mod-friedmann-g,eq:mod-friedmann-f} entirely determine
the dynamics; the spatial components of the Einstein field \cref{eq:einstein-eq}
do not provide further information.

Combining the modified Friedmann \cref{eq:mod-friedmann-g,eq:mod-friedmann-f} yields a quartic polynomial for $y$,
\begin{flalign}\label{eq:quartic-pol-y}
	\alpha^2\beta_3 y^4+(3\alpha^2\beta_2-\beta_4) y^3+3(\alpha^2\beta_1-\beta_3)y^2+\left(\alpha^2\beta_0 -3\beta_2 + \frac{\alpha^2}{m_g^2}\rho_\mathrm{m}\right)y -\beta_1=0\,,
\end{flalign}
which determines $y$ as a function of $\rho_\mathrm{m}$. Taking the derivative w.r.t. 
$e$-folds $\ln a$ and using the continuity \cref{eq:continuity}, we arrive at 
\cite{Akrami:2012vf}
\begin{flalign}\label{eq:y-prime}
	y^\prime = \frac{3 (1+w_\mathrm{m}) \alpha^2 y^2 \rho_\mathrm{m}/m_g^2}{ \beta_1 - 3 \beta_3 y^2 - 2\beta_4 y^3 + 3\alpha^2 y^2 (\beta_1 +2\beta_2 y + \beta_3 y^2) } \,,
\end{flalign}
where prime denotes derivative with respect to $e$-folds\footnote{
The derivative w.r.t. time $t$ and $e$-folds $\ln a$ of a quantity $A$ are related as 
$\dot A = H A^\prime$.}.
$y^\prime$ is only a function of $y$.
The variable $y$ captures the dynamics of the cosmological solutions entirely and has
a one-dimensional phase-space.

Note that 
the exchange symmetry~\eqref{eq:Z2-symmetry} on the level of 
the FLRW background reads
\begin{flalign}
	y \rightarrow -y \,,\ \ \beta_n \rightarrow (-1)^n \beta_n\,.
\end{flalign}
For arbitrary $\beta_n$ we can restrict ourselves to solutions with $y>0$ without loss of generality.

\subsection{Finite and infinite branch}

As mentioned previously, in the presence of a massive spin $2$-field on an (A)dS 
background, the Higuchi bound~\eqref{eq:Higuchi-bound} has to be
satisfied in order to ensure unitarity. Despite the group-theoretical origin of this bound which is only well defined
on (A)dS or Minkowski, it can be generalized to FLRW space~\cite{Fasiello:2013woa}.
Demanding the absence of ghosts results in the cosmological stability bound
\begin{flalign}
	m_\mathrm{eff}^2 \geq 2H^2\,.
\end{flalign}
in terms of the effective mass parameter
\begin{flalign}
	m_\mathrm{eff}^2 = \left( 1 + \frac{1}{\alpha^2 y^2} \right) y (\beta_1 + 2\beta_2 y + \beta_3 y^2)\,.
\end{flalign}
In vacuum with $\rho_\mathrm{m}=0$, i.e. $y= c$ and $H^2= \Lambda/3$, the cosmological stability
bound reduces to the usual Higuchi bound.

We can rewrite \cref{eq:y-prime} in terms of the cosmological stability bound as
\begin{flalign}\label{eq:y-prime-phys}
		y^\prime =  y \frac{(1+w_\mathrm{m}) \rho_\mathrm{m}/m_g^2}{m_\mathrm{eff}^2-2H^2}\,.
\end{flalign}
This allows to read off some important features of cosmological solutions.
First of all, $y^\prime=0$ if $\rho_\mathrm{m}=0$ or $y=0$.
That means, the points $y=0$ and $\rho_\mathrm{m}(y)=0$ cannot be crossed dynamically.
These points separate regions of the phase space of different branches
of solutions to \cref{eq:quartic-pol-y}.
In particular, the vacuum points $\rho_\mathrm{m}(y)=0$ cannot be crossed dynamically.
$y$ approaches a
constant value as can be seen from the~\cref{eq:quartic-pol-y} for
vanishing matter energy density.
From the quartic polynomial~\eqref{eq:quartic-pol-y} we can identify two different
behaviors of $y$ for early times when the matter energy density is large and
classically diverges, $\rho_\mathrm{m}\rightarrow\infty$:
\begin{enumerate}
	\item \textit{Infinite branch:} At early times, $y$ diverges as well. The cosmic evolution starts 
	at $y=\infty$ and as the universe expands, $y$ 
	decreases and finally approaches a constant value $y=c$. This constant 
	corresponds to the highest-lying, strictly positive root of \cref{eq:cquartic}.
	Since $y$ decreases in time, it follows that $y^\prime<0$.
	Now~\cref{eq:y-prime-phys} implies that either $m_\mathrm{eff}^2<2H^2$
	or $\rho_\mathrm{m}<0$.
	Hence, the infinite branch either violates the cosmological stability
	bound or the matter sector has a negative energy density.
	This implies that the infinite branch necessarily propagates a ghost~\cite{Konnig:2015lfa};
	it is an unphysical solution.

	\item \textit{Finite branch:} 
	Alternatively, $y\rightarrow0$ at early times.
	Then $y$ increases in time until it approaches a constant
	value $y=c$ in the infinite future, which is the lowest-lying, strictly positive root
	of \cref{eq:cquartic}. This implies that $y^\prime>0$ and 
	due to~\cref{eq:y-prime-phys} the cosmological stability bound is satisfied.
	This identifies the finite branch as the only solution to the Friedmann~\cref{eq:mod-friedmann}
	that is physical.
	
\end{enumerate} 
Besides the finite and infinite branch, the polynomial \eqref{eq:quartic-pol-y} has up to 
four solutions. These \textit{exotic branches} however were found to not be consistent
\cite{Konnig:2015lfa}.
For a detailed discussion on the viability of cosmological solutions although
in a different parametrization see Ref.~\cite{Konnig:2013gxa}.
From now on, we will only focus on a cosmic expansion history
on the finite branch.

\section{Unique vacuum and physical parametrization of solutions}
\label{sec:vacuum-solutions}

Our aim in this paper is to use cosmological observables to constrain the physical
parameters $\bar\alpha$, $\mFP$ and $\Lambda$. However, these parameters can
be defined only on proportional solutions and hence in vacuum, while cosmological
observables are entities of solutions with matter source.
The idea is to use the asymptotic future of the universe as the vacuum point
at which the spin-$2$ mass $\mFP$, mixing angle $\bar\alpha$ and cosmological
constant $\Lambda$ are defined and to impose consistency conditions on the expansion
history and the asymptotic vacuum point.
This results in a unique relation between the parameters that appear in the action and
the physical parameters.
In this section we work out this strategy in detail and build up the dictionary
between the two different parameterizations.

\subsection{Rescaling invariance and natural parameter values}

The action~\eqref{eq:bimetric-action} has seven free parameters $\{m_g, \alpha, \beta_n\}$.
Due to the properties of the elementary symmetric polynomials
that appear in the bimetric potential~\eqref{eq:pot-interchange-symmetry},
the action is invariant under the combined rescaling
\begin{flalign}\label{eq:rescaling}
	\fmn \rightarrow \tilde f_{\mu\nu} = \lambda^{-1} \fmn \ , \
	\alpha \rightarrow \tilde\alpha = \lambda^{1/2} \alpha \ , \ 
	\beta_n \rightarrow \tilde\beta_n = \lambda^{n/2} \beta_n \,,
\end{flalign}
where $\lambda$ is a constant parameter.
On proportional background solutions, the rescaling of the metric $\fmn$ translates into
\begin{flalign}
	c \rightarrow \tilde c = \lambda^{-1/2} c \,.
\end{flalign}
This implies that one of the eight parameters $\{m_g, \alpha,c,\beta_n\}$ is redundant.
In order to remove the redundancy from the parameter space, the rescaling 
has often been used to either set $\tilde\alpha=1$ 
by choosing $\lambda=\alpha^{-2}$ or to set
$\tilde c=1$ by choosing $\lambda=c^2$ in the literature.
Let us call this choice to fix the redundancy \textit{rescaled parametrization}.
Although being consistent, this choice
leads to a very specific region of the bimetric
parameter space, in which certain features of bimetric theory are not obvious.
This becomes particularly important when studying limits of the theory such as the GR-limit
or the massive gravity limit. 
Suppose, the interaction parameters 
are all of the same order, $\beta_n \sim \mathcal O (m^2)$ where $m$
is some mass scale, e.g. $m=H_0$. Using the rescaling invariance to 
set $\tilde \alpha=1$, the interaction parameters in rescaled parametrization are of the order
$\tilde\beta_n\sim \alpha^{-n}\mathcal O(m^2)$.
In the GR-limit of the theory $\alpha \ll 1$, 
the rescaled interaction parameters are no longer of the same order.
Instead, there is a huge hierarchy between them, 
$\tilde\beta_n\ll \tilde\beta_{n+1}$.
When working in rescaled parametrization one has to impose a large
hierarchy between the interaction parameters in order to
arrive at the GR-limit of bimetric theory.
At first glance, such a parameter choice appears unnatural which led to
confusion in the past on the phenomenological viability of bimetric
theory~\cite{Akrami:2015qga}.

As shown in Ref.~\cite{Babichev:2016bxi}, solutions to the bimetric field
equations exhibit another GR-limit. 
If the massive spin $2$-field is heavy, $\mFP^2\gg\Lambda$, the laws of GR are recovered.
In order to achieve a large Fierz-Pauli mass by keeping the cosmological constant small
requires a large amount of tuning among the interaction parameters $\tilde\beta_n$.
Although this tuning appears to be unnatural, it is another artifact of the rescaling.

To see that, let us briefly discuss the relation between the different parameters
without rescaling. First, quantities like the Fierz-Pauli mass
and the cosmological constant are defined in proportional background
solutions, labeled by the roots $c$ of~\cref{eq:cquartic}.
For a generic model, we can distinguish two types of roots by their asymptotic behavior:
\begin{itemize}
	\item singular root: $c\sim \alpha^{-1}$ as $\alpha\ll1$
	\item constant root: $c$ constant as $\alpha\ll 1$\,.
\end{itemize}
For both types of roots, the Fierz-Pauli mass becomes large in the limit $\alpha\ll1$ 
if we do not tune the $\beta_n$.
On a singular root, however, the cosmological constant is large as well for $\alpha\ll1$.
In order to achieve the hierarchy on a singular root, one has to tune one
of the interaction parameters $\beta_n$.
On a constant root however, the value of $c$ is such that the cosmological constant
is independent of $\alpha$ and of the order of the $\beta_n$ in the limit,
as can be seen from~\cref{eq:eff-cosmological-constant-g}.
Summarizing, $\alpha\ll1$ automatically implies $\mFP^2\gg\Lambda$ without
further tuning the interaction parameters on a constant root.
Alternatively, one can achieve a large Fierz-Pauli mass, $\mFP^2\gg\Lambda$, even 
though $\alpha$ is not small by tuning the interaction parameters $\beta_n$ (and vice
versa). In fact, $\alpha$ and $\mFP$ are completely independent of each other if one
accepts tuning among the interaction parameters\footnote{For the question of
naturalness of such tuning, we refer to
Refs.~\cite{deRham:2013qqa,deRham:2014naa,Heisenberg:2014rka} which
studied the quantum corrections coming from matter and graviton loops that
the bimetric potential recieves.}.
We demonstrate this point for a concrete example in~\cref{sec:tuning}.

\subsection{Definition of physical parameters}

In this paper, we are seeking a parametrization of solutions to the bimetric field
equations that avoids the redundancy due to the rescaling invariance
\eqref{eq:rescaling} and circumvents the aforementioned difficulties that come
along with fixing the redundancy by hand.
Our proposal is to not work in terms of the parameters of the theory $\{\alpha,\beta_n\}$
as independent parameter, but a different set of parameters, that (a) are invariant
under the rescaling~\eqref{eq:rescaling} and (b) have a direct physical
interpretation and capture the relevant limits of bimetric theory.
These independent parameters are:
\begin{itemize}
	\item mixing angle: $\bar\alpha=\alpha c$\,,
	\item Fierz-Pauli mass: $m_\mathrm{FP}$\,,
	\item effective cosmological constant: $\Lambda$\,.
\end{itemize}
Those are the \textit{physical parameters} that can be measured by
local experiments. If there are three free interaction
parameters $\beta_n$, all three physical parameters are independent.
If there are less, the physical parameters are not independent
of each other. For four or five free interaction parameters, we 
additionally introduce the
\begin{itemize}
	\item invariant interaction parameters: $\bar \beta_n = \alpha^{-n} \beta_n$\,.
\end{itemize}
This completes our list of quantities in \textit{physical parametrization}.
We treat the physical parameters as independent
variables and are agnostic to the underlying values of the
parameters of the theory. Instead, the parameters that appear in
the action, are functions of the physical quantities, 
$\alpha=\alpha(\bar\alpha,\mFP,\Lambda,\bar\beta_n)$ 
and $\beta_n = \beta_n(\bar\alpha,\mFP,\Lambda,\bar\beta_n)$.

The physical parametrization comes with its own drawbacks.
The physical parameters are parameters of a particular solution, but they
are not parameters of the theory. The relation between
the theory and physical parameters is ambiguous.
The background~\cref{eq:cquartic} is a polynomial in $c$ of degree $4$.
It has up to four real-valued roots, $c_i$. Each root describes a vacuum
of bimetric theory and hence $c_i$ labels vacua. Each vacuum is characterized
by its own mixing angle $\bar\alpha(c_i)$, spin $2$-mass $\mFP(c_i)$, and
cosmological constant $\Lambda (c_i)$. In other words, for a given set of theory
parameters $\{\alpha,\beta_n\}$ there are up to four different sets of physical parameters.
However, as it turns out there is only a single consistent vacuum out of
the four vacuum solutions. Restricting ourselves to the consistent vacuum
implies a unique relation between the theory and physical
parameters.
We will discuss this point in detail in the following and thereby
define what we mean by consistent vacuum. We work out the dictionary between
theory and physical parameters model by model in \cref{sec:submodels,sec:submodels-dict-appenix}.

\subsection{Consistent vacuum}

Not every vacuum solution is consistent
for a given set of theory parameters $\{\alpha,\beta_n\}$.
First of all, we restrict ourselves to positive roots of~\cref{eq:cquartic},
\begin{flalign}\label{eq:pos-c}
	c>0\,,
\end{flalign}
in order to remove the previously mentioned redundancy from the
solution space.
Then, a consistent vacuum propagates a massive spin-$2$ field with
a positive Fierz-Pauli mass,
\begin{flalign}\label{eq:pos-mfp}
	\mFP>0 \,.
\end{flalign}
We are only interested in vacua with a positive cosmological constant, i.e. in de Sitter vacua,
\begin{flalign}\label{eq:pos-L}
	\Lambda > 0 \,,
\end{flalign}
although this is not a theoretical consistency requirement.
Finally, the Higuchi bound~\cite{Higuchi:1986py,Higuchi:1989gz} must be satisfied,
\begin{flalign}\label{eq:pos-Higuchi}
	\mFP^2 \geq \frac{2}{3}\Lambda\,,
\end{flalign}
for a physical solution.

Besides satisfying these criteria, the physical de Sitter vacuum must be
the asymptotic future of the universe. Since only the finite branch
gives rise to a viable expansion history, we demand that the consistent vacuum corresponds
to the final point of the cosmic evolution along the finite branch.
The scale factor ratio $y$ smoothly evolves from zero in the asymptotic past to a constant value
in the asymptotic future when $\rho_\mathrm{m}\rightarrow 0$.
Hence, the asymptotic future is the lowest-lying,
strictly positive root of~\cref{eq:cquartic}.
This uniquely determines the true vacuum of bimetric theory.
The existence and consistency of the finite branch imposes another
constraint on the theory parameters.
At early times, $y$ approaches zero and the Hubble
rate diverges as the matter energy density diverges.
From the $\fmn$-Friedmann~\cref{eq:mod-friedmann-f} we find
that $H^2\rightarrow \infty$ as $y\rightarrow 0$ for $y>0$ only if~\cite{Konnig:2013gxa}
\begin{flalign}\label{eq:pos-b1}
	\beta_1>0\,.
\end{flalign}
This translates into another consistency constraint on the physical parameters.

Summarizing, for a given set of theory parameters we define a vacuum point to be consistent if
it is the lowest lying, strictly positive root of~\cref{eq:cquartic} 
and satisfies the criteria~\cref{eq:pos-c,eq:pos-mfp,eq:pos-L,eq:pos-Higuchi,eq:pos-b1}.

The previously described procedure identifies the unique consistent vacuum of bimetric theory
for a given set of theory parameters.
Once having identified the consistent vacuum, we
use~\cref{eq:eff-cosmological-constant,eq:fierz-pauli-mass}
to express the theory parameters $\{\alpha,\beta_n\}$ in terms of the 
physical parameters $\{\bar\alpha,m_\mathrm{FP},\Lambda\}$.
It allows to rewrite the Friedmann equation and related cosmological quantities 
in terms of physical parameters, which then can be constrained by cosmological data.
We will work out the dictionary in~\cref{sec:submodels,sec:submodels-dict-appenix} and use it to
constrain the physical parameter space with supernovae data
in~\cref{sec:obs-constraints}.

\subsection{GR- and massive gravity limit in physical parametrization}

Let us briefly comment on the limits in physical parametrization in which either General Relativity
or Massive Gravity is recovered.

As derived in
Refs.~\cite{Akrami:2015qga,Babichev:2016bxi} and applied to concrete examples
in Ref.~\cite{Luben:2018ekw}, bimetric solutions have two independent
parameter regimes in which the laws of GR are recovered:
\begin{enumerate}
	\item $\bar\alpha\ll1$\,,
	\item $\mFP\gg \ell^{-1}$\,,
\end{enumerate}
where $\ell$ is some typical length scale of the system, e.g. $\ell = \Lambda^{-1/2}$.
In the limit of small mixing angle, the fluctuations of the physical metric is
aligned with the massless mode, $\delta \gmn \simeq \delta G_{\mu\nu}$. On the
other hand, the fluctuations of the auxiliary metric is aligned with the massive mode
$\delta \fmn\simeq \delta M_{\mu\nu}$.

The massive gravity limit is arrived at by taking $\bar\alpha \gg 1$ as then
the metric fluctuation is aligned with the massive fluctuations, 
$\delta \gmn \simeq \delta M_{\mu\nu}$. In this limit, the gravitational force between
two test particles is mediated by the massive spin-$2$ field only,
which gives rise to a Yukawa-type gravitational
potential, cf. e.g. Refs.~\cite{Babichev:2013pfa,Comelli:2011wq,Baccetti:2012bk}.
This limit of the parameter space of bimetric theory is highly constrained by observational data
\cite{deRham:2016nuf}.

\section{Model-specific considerations}\label{sec:submodels}

After having established the physical parametrization, we will now explicitly
identify the consistent de Sitter vacuum and the resulting relation
between the theory and physical parameters for various (sub-)models of
bimetric theory. We restrict our analysis to those models
where we can identify the unique consistent vacuum for generic theory
parameters. These are all bimetric models with up to
three non-vanishing interaction parameters.
The reader mostly interested in the constraints from cosmological data
on the physical parameter space might want to jump to the
next section.

Since the polynomial structure of the background
\cref{eq:cquartic} and the quartic \cref{eq:quartic-pol-y} differ from model to model,
we did not find a unified treatment to discuss a generic model.
The following procedure applies to models with at least two non-vanishing interaction parameters;
the model(s) with only one non-vanishing interaction parameter will be discussed separately.
The recipe goes as follows:
\begin{enumerate}
	\item Replace two interaction parameters $\beta_n$ by the physical quantities 
	$\mFP$ and $\Lambda$ using 
	\cref{eq:eff-cosmological-constant,eq:fierz-pauli-mass}.
	\item Solve the background \cref{eq:cquartic} for the quantity 
	$\bar\alpha=\alpha c$ in terms of the physical quantities and the remaining 
	interaction parameters $\bar\beta_n=\alpha^{-n}\beta_n$. Each $\bar\alpha$ represents 
	a vacuum of the model.
	\item Select those vacua $\bar\alpha$ that satisfy the consistency conditions
	in~\cref{eq:pos-c,eq:pos-mfp,eq:pos-L,eq:pos-Higuchi,eq:pos-b1}. 
	For each set of parameters, pick the lowest-lying strictly 
	positive root $\bar\alpha$.
	\item For each consistent vacuum, invert the expression for 
	$\bar\alpha=\bar\alpha (m_\mathrm{FP},\Lambda,\bar\beta_n)$ 
	to express one of the remaining interaction parameters in terms of physical parameters.
	Plug the result into the other expressions.
	\item From the requirement $\beta_1>0$ find further constraints on the physical parameters.
	\item Solve~\cref{eq:quartic-pol-y} for $\rho_\mathrm{m}(y)$ and identify the parameter
	region where $\rho_\mathrm{m}\rightarrow0$ as $\alpha y\rightarrow \bar\alpha$ for
	$\alpha y<\bar\alpha$.
	This ensures that $y'$ does not diverge on the finite branch.
\end{enumerate}
This procedure identifies those parameter ranges in which a viable finite branch exists
with a consistent vacuum as the asymptotic future.
Moreover, it replaces three theory parameters by physical parameters.
Friedmann's equation and all other observables depend only on manifestly
rescaling-invariant parameter combination (e.g. $\beta_n y^n$).

For later use, let us introduce the  parameters
\begin{flalign}\label{eq:pheno-params}
	\Omega_\mathrm{DE} = \frac{\rho_\mathrm{DE}}{3H^2m_g^2} \,,\
	\Omega_\mathrm{m} = \frac{\rho_\mathrm{m}}{3 H^2 m_g^2} \,,\
	\Omega_\Lambda = \frac{\Lambda}{3H^2}\,,\ 
	\Omega_\mathrm{FP} = \frac{m_\mathrm{FP}^2}{3H^2}\,, \
	B_n = \frac{\alpha^{-n}\beta_n}{3H^2}
\end{flalign}
inspired by the standard energy density parameters.
Note that $\Lambda$ is the cosmological constant in the asymptotic future.
The GR-relation $1=\Omega_\Lambda+\Omega_\mathrm{m}$ does not 
hold in bimetric theory.
Instead, the bimetric Friedmann equation can be written as
\begin{flalign}
	1 = \Omega_\mathrm{DE} + \Omega_\mathrm{m}\,.
\end{flalign}
The parameter $\Omega_\mathrm{DE}$ describes the energy density that originates from 
the bimetric potential and is a complicated function 
of time and the other bimetric parameters.
While $\Omega_\Lambda$, $\Omega_\mathrm{FP}$ are time-dependent only via $H$,
the time evolution of the matter energy density is standard,
\begin{flalign}
	\Omega_\mathrm{m}=\frac{H_0^2}{H^2}\,\Omega_\mathrm{m,0} (1+z)^{3(1+w_\mathrm{m})}\,
\end{flalign}
where the redshift $z$ is related to the scale factor of the physical metric $\gmn$ 
in the standard way as $a=(z+1)^{-1}$, cf. \cref{eq:matter-conservation}.
Evaluating the Friedmann equation today yields
\begin{flalign}
	\Omega_\mathrm{m,0}=1-\Omega_\mathrm{DE,0}\,.
\end{flalign}
We can use this relation in order to remove one parameter
from the Friedmann equation. However, the precise relation
between $\Omega_\mathrm{m,0}$, $\Omega_{\Lambda,0}$, $\Omega_\mathrm{FP,0}$, 
$\bar\alpha$, and $B_{n,0}$ depends on the model. 

In the remainder of this section, we will discuss several (sub-)models
of bimetric theory and build up the dictionary
between the theory and physical parameters.
We complete the dictionary for the three parameter models
in~\cref{sec:submodels-dict-appenix}.

\subsection{$1$-parameter model: $\beta_1$-model}

As a warm-up let us first discuss the one parameter models in 
order to demonstrate the procedure.
The only one parameter model that can possibly give rise to a viable
expansion history is the $\beta_1$-model. 
With only one interaction parameter being non-zero, the parameter
space is highly restricted and the physical parameters are not independent. 
Instead, \cref{eq:cquartic,eq:eff-cosmological-constant,eq:fierz-pauli-mass} imply
\begin{flalign}
	\bar\alpha^2=\frac{1}{3}\ , \ \mFP^2=\frac{4}{3}\Lambda\,.
\end{flalign}
The model has two vacua, $\bar\alpha_\pm=\pm\frac{1}{\sqrt{3}}$. Both vacua 
satisfy the Higuchi bound, but only $\bar\alpha_+$ is strictly positive.
The interaction parameter is given by $\alpha^{-1}\beta_1=\Lambda/(3\bar\alpha)$,
which is manifestly positive on the positive vacuum as desired.
Thus, we identified the unique consistent vacuum of the $\beta_1$-model.
Note that $\bar\alpha$ is constant and not a free parameter. Both vacua
of the $\beta_1$-model do not have a GR- or massive gravity limit.

Next, we study the finite branch of the FLRW solution.
The quartic \cref{eq:quartic-pol-y} has two solutions for $y$,
\begin{flalign}\label{eq:b1-y-sol}
	\bar y_\pm = \frac{\pm\sqrt{4\Lambda+\frac{\rho_\mathrm{m}^2}{m_\mathrm{g}^4}}-\frac{\rho_\mathrm{m}}{m_\mathrm{g}^2}}{2\sqrt{3}\Lambda}\,,
\end{flalign}
of which $\bar y_+$ corresponds to the finite branch as can be seen from
the limits of large and small $\rho_\mathrm{m}$.
Now we explicitely see that $\bar y_\pm\rightarrow\bar\alpha_\pm$ in the asymptotic future.
Plugging $y_+$ into the modified Friedmann~\cref{eq:mod-friedmann-g} gives
\begin{flalign}\label{eq:b1-friedmann}
	3H^2 = \frac{\rho_\mathrm{m}}{2 m_g^2} + \sqrt{\Lambda^2 + \frac{\rho_\mathrm{m}^2}{4 m_g^4}}\,.
\end{flalign}
With the Friedmann equation in this parametrization, we can use cosmological observables to constrain the physical paramaters.

Rewriting the modified Friedmann eq. in terms of the parameters in~\eqref{eq:pheno-params} yields
\begin{flalign}
	2 = \Omega_\mathrm{m}+\sqrt{\Omega_\mathrm{m}^2+4\OL^2}
\end{flalign}
Evaluating the Friedmann equation today at redshift $z=0$ yields the relation
\begin{flalign}
	\Omega_{\Lambda,0}=\sqrt{1-\Om}\,.
\end{flalign}
This implies, that the contribution from the bimetric potential today reads
$\Omega_\mathrm{DE,0}=\Omega_{\Lambda,0}^2$.
Now we have all the ingredients to finally compare the model to cosmological data.
This will be done in the next section.

\subsection{$2$-parameter models}

\begin{figure}
\begin{minipage}[c]{0.52\textwidth}
\centering
\includegraphics[scale=0.9]{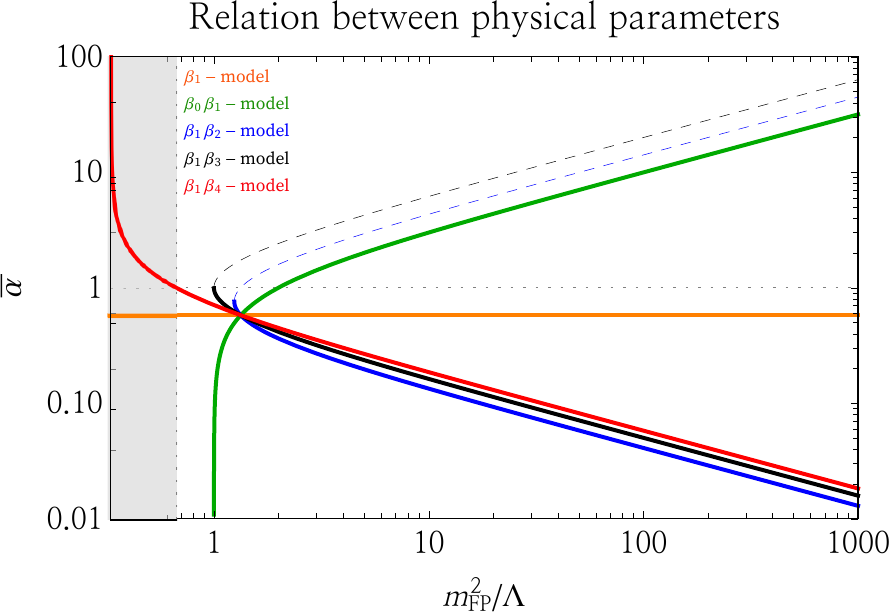}
\end{minipage}
\hfill
\begin{minipage}[c]{0.46\textwidth}
\centering
\begin{tabular}{l | l l}
\hline\hline
Model & GR limit & MG limit \\
\hline
$\beta_1$ & $-$ & $-$ \\
$\beta_0\beta_1$ & $m_\mathrm{FP}^2=\Lambda$ & $m_\mathrm{FP}^2\gg\Lambda$ \\
$\beta_1\beta_2$ & $m_\mathrm{FP}^2\gg\Lambda$ & $-$ \\
$\beta_1\beta_3$ & $m_\mathrm{FP}^2\gg\Lambda$ & $-$ \\
$\beta_1\beta_4$ & $m_\mathrm{FP}^2\gg\Lambda$ & $(m_\mathrm{FP}^2=\Lambda/3)$ \\
\hline\hline
\end{tabular}
\end{minipage}

\begin{minipage}[t]{0.52\textwidth}
\captionof{figure}{This figure shows the relation between $\bar\alpha$ and
$\mFP^2/\Lambda$ for the $1$- and $2$-parameter models on the consistent vacuum.
The dashed lines correspond to the highest-lying, strictly positive root of the model.
In the gray-shaded region the Higuchi bound is violated.}
\label{fig:12-param-models}
\end{minipage}
\hfill
\begin{minipage}[t]{0.46\textwidth}
\captionof{table}{Summary of the general relativity (GR) limit $\bar\alpha\ll1$ and the
massive gravity (MG) limit $\bar\alpha\gg1$ for the one and two parameter models.
A $-$ indicates that the limit is not consistent.}
\label{tab:GR-MG-limits}
\end{minipage}
\end{figure}

Let us discuss models with two interaction parameters being non-zero.
Since $\beta_1$ must be non-zero for the existence of the finite branch, 
we are left with four $2$-parameter models that can possibly give
rise to a viable expansion history: $\beta_0\beta_1$, $\beta_1\beta_2$,
$\beta_1\beta_3$, and $\beta_1\beta_4$.
Only two of the three physical parameters are independent.
In~\cref{fig:12-param-models} we show how $\bar\alpha$ depends
on $\mFP^2/\Lambda$, which we derive in the following subsections.
Note already, that only for the $\beta_0\beta_1$-model, the parameter
$\bar\alpha$ has a range from zero to infinity. For the other
two parameter models, $\bar\alpha<1$ always. This becomes clear when
working out the precise relation between the physical parameters.

The dependency among the physical parameters has important consequences
for the existence of a well-defined GR-limit and massive gravity (MG) limit.
The $\beta_1$-model does not have a GR- or MG-limit at all because $\bar\alpha$
is fixed by the equations of motion to a constant value. For the $\beta_0\beta_1$-model,
the GR-limit is arrived at by taking $\mFP^2=\Lambda$, while the MG-limit
is arrived at by $\mFP^2\gg\Lambda$. For the other two parameter models the situation is
different, where the GR-limit is characterized by $\mFP^2\gg\Lambda$, while they
do not have a consistent MG-limit. Only for the $\beta_1\beta_4$-model, in principle one
can achieve $\bar\alpha\gg1$ by taking $\mFP^2=\Lambda/3$, which however
violates the Higuchi bound. Summarizing, of the $1$- and $2$-parameter models,
only the $\beta_0\beta_1$-model has a consistent massive gravity limit.
This can be seen from~\cref{fig:12-param-models} and the precise relations
that we derive in the following subsections. The GR- and MG-limits are
summarized in~\cref{tab:GR-MG-limits}.

After this summary, let us apply the procedure introduced earlier
in order to express the interaction parameters in terms
of physical parameters.

\subsubsection{$\beta_0\beta_1$ -model}

We start by analysing the model with $\beta_2=\beta_3=\beta_4=0$.
Solving the background equations \cref{eq:cquartic,eq:eff-cosmological-constant,eq:fierz-pauli-mass} yields
\begin{subequations}\label{eq:ParamReps01}
\begin{flalign}
	\bar\alpha_\pm &= \pm\sqrt{\frac{\mFP^2}{\Lambda}-1}\\
	\beta_0 &= -3\mFP^2+4\Lambda\,,\\
	\alpha^{-1}\beta_1 &= \pm\sqrt{(\mFP^2-\Lambda)\Lambda}\,.
\end{flalign}
\end{subequations}
Only in the parameter range
\begin{flalign}
	\mFP^2>\Lambda\,,
\end{flalign}
the vacuum points and $\beta_1$ are real-valued.
This bound is more restrictive than the Higuchi bound.
In the same parameter range, only the vacuum point $\bar\alpha_+$
is strictly positive and hence we discard $\bar\alpha_-$. 
This uniquely identifies $\bar\alpha_+$ as the consistent de Sitter vacuum
of the $\beta_0\beta_1$ model.

Next, we find the roots of \cref{eq:quartic-pol-y} of which only one describes a consistent finite branch.
Plugging the result into \cref{eq:mod-friedmann-g} and using \cref{eq:ParamReps01,eq:pheno-params} yields
the modified Friedmann equation
\begin{flalign}
	2 = -3\OFP+4\OL + \Omega_\mathrm{m} + \sqrt{(3\OFP-2\OL)^2+2(-\OFP+4\OL)\Omega_\mathrm{m}+\Omega_\mathrm{m}^2}\,.
\end{flalign}
Evaluating the Friedmann equation today implies
\begin{flalign}
	\Omega_\mathrm{FP,0} = \Omega_{\Lambda,0}-\frac{1}{3} + \frac{\Om}{3-3\Omega_{\Lambda,0}}\,,
\end{flalign}
which allows to eliminate one of the parameters.
Consistency requires $\Omega_\mathrm{FP,0}>\Omega_{\Lambda,0}$ and
$\Omega_\mathrm{FP,0}>0$, which translates into
\begin{flalign}
	&\Omega_{\Lambda,0} > 1-\Om\,\\
	&\Omega_{\Lambda,0} > \frac{2-\sqrt{1+3\Om}}{3}\,.
\end{flalign}
This completes the dictionary for the $\beta_0\beta_1$ model.

\subsubsection{$\beta_1\beta_2$-model}

Next, we consider the model with $\beta_0=\beta_3=\beta_4=0$.
In physical parametrization, the background \cref{eq:cquartic} has four roots. 
The two solutions with $\bar\alpha>0$ are
\begin{subequations}\label{eq:ParamReps12}
\begin{flalign}
	\bar\alpha_\pm &= \sqrt{\frac{3\mFP^2 - 2\Lambda \pm \sqrt{9\mFP^4 -12\mFP^2\Lambda + \Lambda^2}}{3\Lambda}}\,,\\
	\alpha^{-1} \beta_{1\pm} &= \frac{\bar\alpha_\pm}{2} \left( 3\mFP^2 - 3\Lambda \mp \sqrt{9\mFP^4 -12\mFP^2\Lambda + \Lambda^2} \right)\,,\\
	\alpha^{-2} \beta_{2\pm} &= - \frac{1}{6} \left( 3\mFP^2 - 5\Lambda \mp \sqrt{9\mFP^4 -12\mFP^2\Lambda + \Lambda^2} \right)\,.
\end{flalign}
\end{subequations}
The other two roots are given by $-\bar\alpha_\pm$, but we discard them due to our 
requirement $\bar\alpha>0$. Both, $\bar\alpha_\pm$ and the 
interaction parameters are real-valued and positive only in the parameter range
\begin{flalign}
	\mFP^2 > \frac{2+\sqrt{3}}{3} \Lambda \,,
\end{flalign}
which is more restrictive than the Higuchi bound.
In the same parameter range, we find that $\bar\alpha_- < \bar\alpha_+$.
This identifies $\bar\alpha_-$ is the lowest-lying strictly positive root and
thus the unique consistent vacuum of the $\beta_1\beta_2$ model.
Therefore, we use $\beta_{1-}$ and $\beta_{2-}$ in order to replace interaction
parameters by physical parameters.

The polynomial~\eqref{eq:quartic-pol-y} has degree $3$ and hence 
there are up to three real-valued roots $y$.
The explicit expressions are lengthy and not enlightening so we do not write them here.
Instead, we only report the result from evaluating the Friedmann equation today.
The relation between the parameters reads
\begin{flalign}
	\Om =& \frac{(\Omega_{\Lambda,0} - 1)}{(2+6\OFPN-10\Omega_{\Lambda,0} - 9\OFPN \Omega_{\Lambda,0} + 12 \Omega_{\Lambda,0} ^2)^2} \Bigg[ -27\OFPN^3 \Omega_{\Lambda,0} \nonumber\\
	+& 9\OFPN^2(4+17\Omega_{\Lambda,0}^2 +\Omega_{\Lambda,0} ( -12+\sqrt{9\OFPN^2 - 12\OFPN \Omega_{\Lambda,0} + \Omega_{\Lambda,0}^2} )) \nonumber\\
	-& 6 \OFPN ( -4 + 24 \Omega_{\Lambda,0} + 43 \Omega_{\Lambda,0}^3 + \Omega_{\Lambda,0}^2 (-50 + 3 \sqrt{9\OFPN^2-12\OFPN \Omega_{\Lambda,0} + \Omega_{\Lambda,0}^2}) ) \nonumber\\
	+& 2 \Big( 2+68\Omega_{\Lambda,0}^4 -\Omega_{\Lambda,0} (18+\sqrt{9\OFPN^2 - 12\OFPN\Omega_{\Lambda,0}+\Omega_{\Lambda,0}^2}) + \Omega_{\Lambda,0}^2 (63+\sqrt{9\OFPN^2-12\OFPN\Omega_{\Lambda,0}+\Omega_{\Lambda,0}^2})\nonumber\\
	+&\Omega_{\Lambda,0}^3(-103 + 4 \sqrt{9\OFPN^2-12\OFPN\Omega_{\Lambda,0}+\Omega_{\Lambda,0}^2}) \Big)\Bigg]
\end{flalign}
These are all the ingredients that we need for the data analysis.

\subsubsection{$\beta_1\beta_3$-model}

The $\beta_1\beta_3$ model is defined by $\beta_0=\beta_2=\beta_4=0$.
The background~\cref{eq:cquartic,eq:eff-cosmological-constant,eq:fierz-pauli-mass}
have the following solutions,
\begin{subequations}\label{eq:ParamReps13}
\begin{flalign}
	\bar\alpha_\pm &= \sqrt{ \frac{2 \mFP^2-\Lambda \pm 2\mFP\sqrt{\mFP^2-\Lambda}}{\Lambda} }\,,\\
	\alpha^{-1} \beta_{1\pm} & = \frac{\bar\alpha^\pm}{4}\sqrt{ 3\mFP^2 -2\Lambda \mp 3\mFP \sqrt{\mFP^2-\Lambda} }\,,\\
	\alpha^{-3} \beta_{3\pm} & = -\bar\alpha^\pm \left( 4\mFP^4 - 7\mFP^2\Lambda + 2\Lambda^2 \mp\sqrt{\mFP^2-\Lambda}(4\mFP^2-5\Lambda)\mFP \right)\,.
\end{flalign}
\end{subequations}
We find that $\bar\alpha_\pm$, $\beta_{1\pm}$, and $\beta_{3\pm}$ are
positive and real-valued only in the parameter range
\footnote{Strictly speaking, $\beta_{1+}$ is positive and real-valued only in the parameter range $\frac{3}{4}\mFP^2<\Lambda<\mFP^2$.
However, $\bar\alpha_+$ is not a consistent vacuum point anyways.}
\begin{flalign}
	\mFP^2>\Lambda \,,
\end{flalign}
which is more restrictive than the Higuchi bound.
In the same parameter region we find that $\bar\alpha_- < \bar\alpha_+$.
This identifies $\bar\alpha_-$ as the unique consistent vacuum.

Evaluating the Friedmann equation for $z=0$ implies 
the following relation among the energy density parameters,
\begin{flalign}
	\Omega_\mathrm{m,0} = &\frac{2\OLN^2}{27( -2\OFPN+\OLN +2\sqrt{\OFPN(\OFPN+\OLN)} )}
	\Bigg[ -108\OFPN -108\sqrt{\OFPN^5(\OFPN-\OLN)} + 297 \OFPN^2\OLN \nonumber\\
	+& 243\sqrt{\OFPN^3(\OFPN-\OLN)}\OLN -243\OFPN\OLN^2 + 2\OLN\left( 2+9\OLN\left(2\OLN-7\sqrt{\OFPN(\OFPN-\OLN)}\right) \right)\nonumber\\
	+&\OLN \left( -2+18\OFPN^2+18\sqrt{\OFPN^3(\OFPN-\OLN)}-33\OFPN\OLN+12\OLN\left(\OLN-2\sqrt{\OFPN(\OFPN-\OLN)}\right) \right)\nonumber\\
	\times& \sqrt{4+18\OFPN^2+18\sqrt{\OFPN^3(\OFPN-\OLN)}-33\OFPN\OLN+12\OLN\left(\OLN-2\sqrt{\OFPN(\OFPN-\OLN)}\right)}
	\Bigg]\,,
\end{flalign}
which agrees with the relation given in Ref. \cite{Konnig:2013gxa}, as 
we checked explicitly.
Since the expressions are to lengthy, we do not show the
Friedmann equation on the finite branch in full glory.

\subsubsection{$\beta_1\beta_4$-model}

Finally, we discuss the $\beta_1\beta_4$-model which is defined by setting
$\beta_0=\beta_2=\beta_3=0$.
The background~\cref{eq:cquartic,eq:eff-cosmological-constant,eq:fierz-pauli-mass}
have the following roots,
\begin{subequations}\label{eq:ParamReps14}
\begin{flalign}
	\bar\alpha_\pm &= \pm \sqrt{\frac{\Lambda}{3\mFP^2-\Lambda}}\,,\\
	 \alpha^{-1}\beta_{1\pm} & = \frac{1}{3} \sqrt{(3\mFP^2-\Lambda)\Lambda}\,,\\
	\alpha^{-4} \beta_{4\pm} & = - \frac{9\mFP^4 - 15\mFP^2\Lambda + 4\Lambda^2}{3\Lambda}\,.
\end{flalign}
\end{subequations}
The roots are real-valued in the parameter range $3\mFP^2>\Lambda$, which
is less restrictive than the Higuchi bound.
In the consistent parameter range, only the root $\bar\alpha_+$ is strictly positive.
This identifies $\bar\alpha_+$ as the unique consistent vacuum of the $\beta_1\beta_4$ model.
When the Higuchi bound is satisfied, also the interaction parameter $\beta_{1+}$ is positive.

Instead of presenting the lengthy expression for the Friedmann equation 
on the finite branch, we only evaluate it today.
The resulting relation among the parameters is
\begin{flalign}
	\OFPN=\frac{\OLN( -4+12\Om -12\Om^2+4\Om^3+3\OLN-\Om \OLN+\OLN^3 )}{3(-1+3\Om-3\Om^2+\Om^3+\OLN^3)}\,.
\end{flalign}
This completes the dictionary of the $2$-parameter models.

\subsection{$3$-parameter models}

In this section we will discuss models with three non-vanishing interaction parameters.
That means that all three physical parameters $\bar\alpha$, $m_\mathrm{FP}$,
and $\Lambda$ are independent
and not fixed by the background equations.
We focus on the two extreme cases:
the three parameter model without vacuum energy ($\beta_0=\beta_4=0$)
and the model with vacuum energy in both sectors ($\beta_2=\beta_3=0$).
We complete the dictionary for the other three parameter models
in~\cref{sec:submodels-dict-appenix}.

\subsubsection{$\beta_1\beta_2\beta_3$-model}

Following the procedure, there are two positive vacuum points $\bar\alpha_\pm$.
The quartic polynomial~\eqref{eq:cquartic} is solved most easily leaving $\beta_2$ as a free parameter.
Then the roots take the form
\begin{flalign}
	\bar\alpha^2_\pm = \frac{2\mFP^2-\Lambda-\bar\beta_2 \pm \sqrt{ (2\mFP^2 - \bar\beta_2)^2-4\mFP^2\Lambda }}{2\bar\beta_2 +\Lambda}\,,
\end{flalign}
where $\bar\beta_2=\alpha^{-2}\beta_2$.
The roots $\bar\alpha_\pm$ are real-valued in the following parameter ranges:
\begin{subequations}\label{eq:b1b2b3-con-vacuum-cond}
\begin{flalign}
	& \bar\alpha_+^2>0 \ \ \Longleftrightarrow\ \ -\frac{\Lambda}{2} < \bar\beta_2 < 2 \mFP (\mFP-\sqrt{\Lambda}),\\
	& \bar\alpha_-^2>0 \ \ \Longleftrightarrow\ \ \bar\beta_2 < 2 \mFP (\mFP-\sqrt{\Lambda})
\end{flalign}
\end{subequations}
We find that $\bar\alpha_- < \bar\alpha_+$ in the parameter range, where
both roots are real-valued.
This identifies $\bar\alpha_-$ is the lowest-lying, strictly positive root
and hence as the unique consistent vacuum of the $\beta_1\beta_2\beta_3$ model.
For $\bar\beta_2\rightarrow -\Lambda/2$ we find that $\bar\alpha_-\rightarrow\infty$ such that we have to exclude this point.

\begin{figure}
	\centering
	\includegraphics[scale=1]{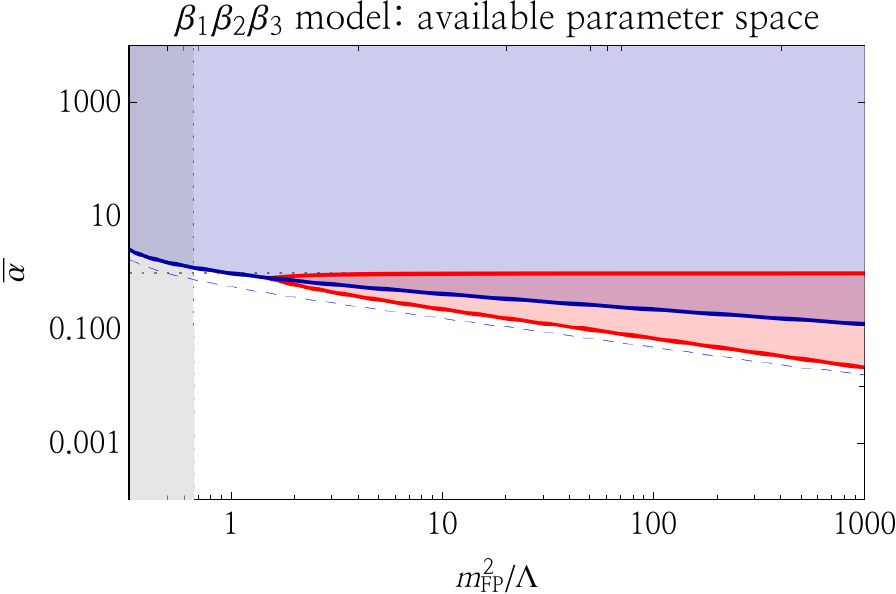}
	\caption{The allowed parameter space in the $\bar\alpha - \mFP$-plane for 
	the $\beta_1\beta_2\beta_3$-model. The spin-$2$ mass is given as
	multiples of the cosmological constant $\Lambda$. Within the red-shaded 
	region, the Hubble rate and the energy density are negative at early times
	because $\beta_1<0$, cf. \cref{eq:b1b2b3-allowed-parameter-space}. 
	In the blue-shaded region, the vacuum point is not well-defined and hence
	this parameter space is excluded as well, cf. 
	\cref{eq:b1b2b3-b2-constraints}. The dashed blue
	line must be excluded, because the parameter replacements 
	are not well-defined, cf. \cref{eq:b1b2b3-b2-constraints}. A viable finite branch
	can only exist outside the colored regions.
	The strongest bounds are
	summarized in \cref{eq:b1b2b3-summary-conditions}.
	Within the gray-shaded region the Higuchi bound is violated.}
	\label{fig:b1b2b3-allowed-parameter-space}
\end{figure}

Solving the expression for $\bar\alpha_-$ for $\beta_2$ and suppressing the label from now on,
we find for the interaction parameters in terms of physical parameters
\begin{subequations}
\begin{flalign}
	\alpha^{-1} \beta_1 & = \frac{ -6 \bar\alpha^2 \mFP^2 + (3+4\bar\alpha^2 + \bar\alpha^4)\Lambda }{4 \bar\alpha ( 1+\bar\alpha^2 )} \,,\\
	\alpha^{-2} \beta_2 & = \frac{ 4\bar\alpha^2 \mFP^2- (1+\bar\alpha^2)^2\Lambda }{2\bar\alpha^2 (1+\bar\alpha^2) } \,,\\
	\alpha^{-3} \beta_3 & = \frac{ (1+3\bar\alpha^4)\Lambda - \bar\alpha^2( 6\mFP^2-4\Lambda )}{ 4\bar\alpha^3 (1+\bar\alpha^2) }\,,
\end{flalign}
\end{subequations}
where the simplified expressions for $\beta_1$ and $\beta_3$ are only valid in the parameter range \cref{eq:b1b2b3-finite-branch}. 
The constraints on the interaction parameter $\beta_2$ translate as follows,
\begin{subequations}\label{eq:b1b2b3-b2-constraints}
\begin{flalign}
	\bar\beta_2 \neq - \frac{\Lambda}{2} \ \ &\Longrightarrow \ \ \mFP^2 \neq \frac{(1+\bar\alpha^2)\Lambda}{4\bar\alpha^2}\,,\\
	\bar\beta_2 < 2 \mFP ( \mFP - \sqrt{\Lambda} ) \ \ & \Longrightarrow \ \ \mFP^2 < \frac{(1+\bar\alpha^2)^2\Lambda}{4\bar\alpha^4}\,.\label{eq:b1b2b3-finite-branch}
\end{flalign}
\end{subequations}
Outside these parameter regions the vacuum $\bar\alpha_-$ is not well-defined
and they have to be excluded from the parameter space
\footnote{Strictly speaking, the condition \eqref{eq:b1b2b3-con-vacuum-cond} translates into 
$4\bar\alpha^4\mFP^2 \neq (1+\bar\alpha^2)^2\Lambda$.
However, we find that for $4\bar\alpha^4\mFP^2 > (1+\bar\alpha^2)^2\Lambda$, the 
expression for the vacuum point is not invertible. We explicitly
checked that in this parameter range, $\rho$ does not vanish at $y=\bar\alpha_-/\alpha$.
Therefore, we have to exclude this parameter region and
this is indicated by the $\Longrightarrow$ in \cref{eq:b1b2b3-b2-constraints}.}.
In \cref{fig:b1b2b3-allowed-parameter-space}, the first bound is represented by the blue-dashed line,
while the second bound is indicated by the blue-shaded region.
On the vacuum point $\bar\alpha_-$, $\beta_1>0$ is satisfied in the parameter ranges
\footnote{Note that region (1) is already excluded by \cref{eq:b1b2b3-b2-constraints}, but we mention it anyways for completness.}
\begin{subequations}\label{eq:b1b2b3-allowed-parameter-space}
\begin{flalign}
	& (1)\ \ \mFP^2 > \frac{(3+2\bar\alpha^2-\bar\alpha^4)}{12(1-\bar\alpha^2)}\Lambda\ \ \mathrm{for} \ \  \bar\alpha \geq \frac{\sqrt{\sqrt{33}-3}}{2}\,,\\
	& (2)\ \ \mFP^2 < \frac{(3+4\bar\alpha^2+\bar\alpha^4)}{6 \bar\alpha^2}\Lambda \ \ \mathrm{for}  \ \ \bar\alpha < \frac{\sqrt{\sqrt{33}-3}}{2}\,.
\end{flalign}
\end{subequations}
In \cref{fig:b1b2b3-allowed-parameter-space} the red-shaded region indicates where $\beta_1<0$.
Moving to cosmology, we can expand the quartic polynomial in \cref{eq:quartic-pol-y}
around $y=\bar\alpha_-$. We find that $\rho_\mathrm{m}$
approaches zero only in the parameter range~\eqref{eq:b1b2b3-finite-branch}.

Let us summarize the most stringent bounds.
The $\beta_1\beta_2\beta_3$-model can give rise to a consistent
expansion history only in the parameter region
\begin{subequations}\label{eq:b1b2b3-summary-conditions}
\begin{flalign}
	& (1) \ \ \frac{\mFP^2}{\Lambda} \leq \frac{(1+\bar\alpha^2)^2}{4\bar\alpha^4} \ \ \ \ \ \ \ \ \ \text{for}\ \ \bar\alpha^2 > \frac{\sqrt{33}-3}{2}\,,\\
	& (2) \ \ \frac{\mFP^2}{\Lambda} < \frac{(3+4\bar\alpha^2+\bar\alpha^4)}{6 \bar\alpha^2} \ \ \text{for}\ \ \bar\alpha^2 < \frac{\sqrt{33}-3}{2}\,,\\
	& (3) \ \ \frac{\mFP^2}{\Lambda} \neq \frac{1+\bar\alpha^2}{4\bar\alpha^2} \ \ \ \ \ \ \ \ \ \ \ \ \ \text{for any } \bar\alpha\,.
\end{flalign}
\end{subequations}
At the point $\bar\alpha^2 = \frac{\sqrt{33}-3}{2}$, the bounds in $(1)$ and $(2)$ coincide with 
$\mFP^2/\Lambda < (19+3\sqrt{33})/24 \approx 1.5 $.
The physical parameter space for the three parameter model without bare cosmological constants
is highly constrained by theoretical consistency requirements. A large Fierz-Pauli mass is
only consistent, if $\bar\alpha$ is sufficiently small. Otherwise the vacuum is not consistent.
Finally, for large spin $2$-masses, the consistency bounds can be summarized
as $\bar\alpha^2<\Lambda/(2\mFP^2)$.

\subsubsection{$\beta_0\beta_1\beta_4$-model}

After having studied the three parameter model without any bare cosmological constant,
we move to the opposite case including vacuum energy for both metrics,
the $\beta_0\beta_1\beta_4$-model.
In physical parametrization but leaving $\beta_0$ free, this model has two roots,
one of which is positive.
It is given by
\begin{flalign}
	\bar\alpha^2 = \frac{\Lambda-\beta_0}{3\mFP^2-\Lambda +\beta_0}
\end{flalign}
and well-defined within the parameter range, where $0<\Lambda-\beta_0<3\mFP^2$.
This is the unique consistent de Sitter vacuum of the $\beta_0\beta_1\beta_4$ model.
Solving this relation for $\beta_0$, we find the interaction parameters in terms of the physical parameters
\begin{subequations}\label{eq:ParamReps014}
\begin{flalign}
	\beta_0 = & \frac{\Lambda - \bar\alpha^2 ( 3\mFP^2 -\Lambda )}{1+\bar\alpha^2}\,,\\
	\alpha^{-1}\beta_1 = & \frac{\bar\alpha \mFP^2}{1+\bar\alpha^2}\,,\\
	\alpha^{-4}\beta_4 = & - \frac{ \mFP^2 - (1+\bar\alpha^2)\Lambda}{\bar\alpha^2 (1+\bar\alpha^2)} \,.
\end{flalign}
\end{subequations}
This expression for $\beta_0$ automatically satisfies the parameter bound
above, when the Higuchi bound is satisfied. Also $\beta_1$ is manifestly
positive. 
Hence, the replacements above are unique and well-defined for the entire 
parameter space in which the Higuchi bound is satisfied.

When expanding the quartic polynomial in \cref{eq:quartic-pol-y} around 
$y=\bar\alpha$ and $y=0$, we find that the Hubble rate and the energy density 
are well-behaved in the same parameter region where the physical parameters 
satisfy the Higuchi bound.

Summarizing, the $\beta_0\beta_1\beta_4$-model has a well-defined and 
consistent expansion history and asymptotic de Sitter point as soon as the 
Higuchi bound is satisfied.
This is in stark contrast to the $\beta_1\beta_2\beta_3$-model where the 
parameter-space is highly restricted by demanding consistency already.
This suggests that the other three parameter models interpolate between
these two extreme cases.

\section{Constraints from SN1a}\label{sec:obs-constraints}

In the previous sections we have collected all the ingredients
allowing us to finally compare the bimetric (sub)models to real data.
We first introduce the observable that provides the data (supernovae)
and explain the data analysis.
In the last subsection we will summarize and discuss the results.

\subsection{Supernovae Type 1a}

In this paper, we focus on supernovae of type 1a as cosmological observable.
Their luminosity can be determined
independently of the redshift which allows to reconstruct the redshift-distance
relationship (the Hubble diagram).
In 1998, supernovae provided the first evidence
that our Universe is currently in a phase of accelerated
expansion~\cite{Riess:1998cb,Perlmutter:1998hx}. Since then, they have become
a powerful tool in constraining cosmological parameters of gravitational theories.
A gravitational theory like bimetric theory predicts
the Hubble rate as a function of redshift $z$, $H=H(z)$. This allows to calculate
various cosmic distances, of which the luminosity distance $d_{\rm L}$ is the relevant
one for supernovae.
In terms of the rescaled Hubble parameter $E(z)=H(z)/H_0$ (where $H_0=H(z=0)$),
the luminosity distance reads
\begin{flalign}\label{eq:luminosity-distance}
	d_{\rm L} (z) = \frac{c}{H_0} (1+z) \int_0^z\frac{\dd z'}{E(z')}\,.
\end{flalign}
Let us already define the rescaled luminosity distance $D_{\rm L}=H_0 d_{\rm L}$,
which is independent of the value of $H_0$.
The apparent magnitude is related to the luminosity distance as
\begin{flalign}
	m = M + 25 + 5 \log_{10} d_{\rm L}
\end{flalign}
where $M$ is the absolut magnitude.
This quantity can be compared to the measured magnitude of a supernova.

The data set we are using is the Union$2.1$ compilation of SN1a
as reported in Ref.~\cite{Suzuki:2011hu}. It contains $580$ supernovae
with redshifts up to $z \lesssim 1.4$.

\subsection{Data analysis}

We aim at constraining the model with real data and thereby finding the regions 
of the physical parameter space of bimetric theory that is in agreement with observations.
We take the Bayesian perspective as it has become standard in cosmology
for parameter estimation.
Let $x$ be a vector of real data and $\theta$ a parameter vector.
Bayes inference relies on Bayes' theorem~\cite{Bayes:1764vd}
\begin{flalign}
	p(\theta | x) = \frac{p (x | \theta) p(\theta)}{p(x)}\,.
\end{flalign}
In observational cosmology the term $p(x)$ is referred to as evidence and
is the probability for the observed data $x$ to occur. The evidence appears as an
overall normalization and is relevant mostly for model comparison
which is not of interest for us in this paper.
In order to find the region of the parameter space that is preferred by the
data, we have to determine the posterior probability distribution function (PDF),
$p(\theta | x)$. It describes
the probability for the parameters $\theta$ to be the true values given the measured data $x$.
We have to choose priors $p(\theta)$ and calculate the likelihood 
$\mathcal L(\theta)\equiv p( x|\theta)$ which we will describe in the next two sections.
Instead of discussing Bayesian inference in further detail, we refer the interested reader to
Refs.~\cite{DAgostini:1995jqe,Trotta:2005ar,Trotta:2008qt,Liddle:2009xe}.

Within Bayesian statistics there are several methods to scan the parameter space
in order to map the posterior PDF. The most straightforward way is to discretize the
parameter space and perform a grid scan. On each point of the grid one calculates
the likelihood $\mathcal L(\theta)$. However, when implemented numerically
this method might be quite slow because computing time is used to evaluate the likelihood
in regions of the parameter space where the model does not give a good fit anyways.
In addition, we are usually interested in the posterior PDF on a subset of the 
full parameter space. This marginalization requires integrating
the posterior PDF over some directions of the parameter space which might
need a lot of computing time. Both drawbacks become particularly relevant when
the parameter space is highly dimensional.
A more efficient approach is the Markov Chain Monte Carlo (MCMC) method
which has become very popular in cosmological data 
analysis~\cite{Christensen:2000ji,Christensen:2001gj}.
Of particular interest is the Metropolis-Hastings algorithm which we choose to use
for our statistical analysis. It explores the parameter
space in regions where the likelihood is large in great detail. It 
automatically yields the marginal posterior PDFs for each parameter individually
as point frequencies of the chains.
To ensure that the result of the algorithm maps the posterior PDFs
to high precision, i.e. ensuring the Markov chains converge,
they have to be long enough. Convergence can be checked, e.g., by comparing
several Markov chains that started in different
regions of the parameter space. For a chain to be independent of the
starting point, a certain amount of steps needs to be removed as a burn-in.
The number of chains, length of each chain and number of burn-in steps
per chain are a matter of choice. Convergence can only be assessed
\textit{a posteriori}.
The Gelman-Rubin factor $\mathcal R$~\cite{Gelman:1992zz} provides some 
quantitative measure of convergence.
For more details on MCMCs and their application in cosmology we refer to
Refs.~\cite{Lewis:2002ah,Gamerman:2006book}.

\subsubsection{Choice of priors}

A key feature in Bayesian inference is the freedom to choose priors $p(\theta)$.
Here we can use all the knowledge about the physical parameter space that 
we have gained in the past. This is the subject of the present section.

Let us start by explicitly stating which are the parameters that we fix
to a certain value although most generally they should be subject to the
statistical analysis. In that sense, strictly speaking these are not priors.
We are interested in times after matter-radiation-equality. Therefore, we
assume the matter energy density to be composed of non-relativistic matter
(such as baryonic and dark matter) only. We set the energy density of
radiation (photons, neutrinos) to zero and the matter equation of state
is consequently $w_\mathrm{m}=0$. In addition, we assume the universe to be exactly
spatially flat although this conclusion was drawn only in the context of the
$\Lambda$CDM model. 
Both these assumptions base their justification on the $\Lambda$CDM model.
Since this model describes the cosmological data to a high precision
and since we expect our models to not deviate to much from
the predictions of the $\Lambda$CDM model (at least at late times),
we believe both assumptions to be justified also within bimetric theory.

Let us move to actual priors. Of course, the energy density parameters
should be positive, $\Om>0$ and $\Omega_{\Lambda,0}>0$, as in $\Lambda$CDM.
In addition, we have defined and derived various consistency conditions
for a model to be viable in~\cref{sec:vacuum-solutions,sec:submodels}.
We use these conditions as priors and accept only those parameter
combinations that satisfy all aforementioned conditions. If any of the
consistency conditions is violated, we set the prior probability for
that parameter combination to zero, and to one otherwise.

These consistency requirements still allow for an infinitely extended
parameter space. That is, $\bar\alpha$ can range from $0$ to infinity.
Also the Fierz-Pauli mass can be arbitrarily large. However, to be physically
meaningful it is limited by the Planck mass.
In order to keep the problem under numeric control, we decided to choose
more restrictive priors. At the same time we are interested in many different
orders of magnitude. Therefore it is natural to work in a logarithmic scale.
Explicitly, we use the uninformative priors
\begin{flalign}
	p(\bar\alpha) = \begin{cases} 1\ ,\ \ -100<\log_{10}(\bar\alpha)<2 \\ 0\ , \ \ \mathrm{else}\end{cases}, \ \
	p(\Omega_{\rm FP,0}) = \begin{cases} 1\ ,\ \ -2<\log_{10}(\Omega_{\rm FP,0})<100 \\ 0\ , \ \ \mathrm{else}\end{cases}.
\end{flalign}
We should point out that the upper limit on $\bar\alpha$ implies that we do not
expect the massive gravity limit of our models to be cosmologically
viable~\cite{DAmico:2011eto,Gumrukcuoglu:2011ew,Gumrukcuoglu:2011zh,Vakili:2012tm,DeFelice:2012mx,Fasiello:2012rw,DeFelice:2013awa}. The lower limit on $\OFP$ is justified
because we expect the value of the effective cosmological constant to be far away enough from
zero for the Higuchi bound to be relevant. The Fierz-Pauli mass cannot be arbitrary small.
The upper bound does not correspond to the Planck scale but rather
$\mFP \simeq 10^{19}\, \mathrm{eV}$, which is of the order of inflation scale. This limit
ensures numerical stability of the evaluations as we checked by various examples.

\subsubsection{Calculating the Hubble rate and likelihood}\label{sec:numerical-investigation}

Now that we clarified where we want to compare the model to observations,
let us explain how. Calculating the luminosity distance as theoretical
prediction, requires integrating
the Hubble rate over redshift $z$, see~\cref{eq:luminosity-distance}. 
The redshift enters the Hubble rate via the matter energy density
$\rho_\mathrm{m}=\rho_\mathrm{m}(z)$, cf.~\cref{eq:matter-conservation}, and 
via the scale factor ratio $y=y(z)$, cf.~\cref{eq:quartic-pol-y}.
Since the bimetric models vary a lot in complexity, we use two different
methods to construct $H(z)$.

For the simple $\beta_1$- and $\beta_0\beta_1$-models, Friedmann's
equation is still handy. As described in the previous section we solved 
\cref{eq:quartic-pol-y} for $y$ and picked the solution that describes
the finite branch. Plugging the result into~\cref{eq:mod-friedmann-g}
yields the Hubble rate as a function of redshift $z$
and we can perform the integration over $z$ numerically in order to get
the luminosity distance $d_{\rm L}$.
It is automatically guaranteed that the expansion history follows
the finite branch.

For all the other models, this procedure turns out to be quite cumbersome and
numerically slow.
Instead, we employ the following strategy.
We solve~\cref{eq:quartic-pol-y} numerically for $y$ at a given
redshift $z$. This yields up to four solutions, out of which we pick the one
that satisfies\footnote{In the numerical analysis we also rescaled the scale factor ratio by
$\alpha$ as $\bar y=\alpha y$ in analogy to $\bar\alpha$. Then Friedmann's equation
and the quartic polynomial are completely independent of $\alpha$.}
$0<\bar y(z)<\bar\alpha$
to ensure that $y$ evolves on the finite branch.
With the resulting value for $y(z)$ we compute the value of the Hubble rate $H(z)$ at
redshift $z$.

In contrast to the analysis of
Ref.~\cite{Akrami:2012vf}, we do not solve the differential equation that
describes the evolution of $y$, cf.~\cref{eq:y-prime}.
Instead, we compute the value for $y$ and hence $H$ for each redshift individually.
Although this requires more computing time, there is no ambiguity in choosing
initial conditions and $y$ is guaranteed to evolve on the finite branch.

Having clarified how to compute the theoretical predictions, let us demonstrate how
to calculate the likelihood.
The absolute magnitude $M$ of a supernova is degenerate with the value of the Hubble rate today, $H_0$,
as these appear as additive quantities\footnote{
To see this, replace $d_{\rm L}$ by $D_{\rm L}$ which yields $m=\mathcal M +5\log_{10}D_{\rm L}$,
where $\mathcal M=M+25-5\log_{10}H_0$ implying that $H_0$ and $M$ are degenerate parameters.
In our analysis, we are not interested in the value of the astrophysical parameter $M$. Therefore,
we marginalize over the parameter $\mathcal M$. Note that in this sense, supernovae do
not provide constraints on $H_0$.}
in $m$.
In order to remove the degeneracy, one can define a new variable which is the sum of both.
This new variable however is not of interest for us in our cosmological data analysis
and appears as a nuisance parameter. We wish to marginalize over the nuisance parameter.
This can be done analytically by defining a marginalized $\chi^2$ as
\begin{flalign}
	\tilde\chi^2(\theta) = \sum_i \frac{(5\log_{10}D_{\rm L}(\theta)-m(x_i))^2}{\sigma_i^2} - \frac{\left(\sum_i\frac{5\log_{10}D_{\rm L}(\theta)-m(x_i)}{\sigma_i^2}\right)^2}{\sum_i \sigma_i^{-2}}\,,
\end{flalign}
where $D_{\rm L}$ is the rescaled luminosity distance which an be computed from $E(z)$.
In each step of the Markov chain we compute the quantity $\tilde\chi^2$.
The likelihood is then given by
\begin{flalign}
	\mathcal L (\theta) = e^{-\frac{1}{2}\tilde\chi^2(\theta)}\,.
\end{flalign}
Having the minimum of the $\chi^2$-distribution for each model as a measure
for the goodness of fit at hand, we would like to compare the different models.
As a rough estimate for model comparison, we introduce the
reduced $\chi^2$ as
\begin{flalign}
	\tilde \chi^2_{\rm red} = \frac{\tilde\chi^2_{\rm min}}{\rm d.o.f.}
\end{flalign}
where the number of degrees of freedom is given by ${\rm d.o.f.}=N-P$
in terms of the number of data points $N$ and number of free parameters of the model $P$.
In our case, we have $N=580$ data points while the number of free parameters varies
from model to model.
Determining the correct number of effective free parameters of a model
is not straightforward, in particular for nonlinear models or correlated
parameters~\cite{Andrae:2010gh}.
As we only want to give a rough estimate, we take $P$ to be
the number of free fitting parameters.
With this value for $P$ we tend to overestimate the value of $\tilde\chi^2_{\rm red}$.

\subsection{Results and discussion}\label{sec:results}

In this section we discuss and summarize the results
of the statistical analysis. The best-fit values for the physical parameters are
summarized in~\cref{tab:best-fit-values} for all models under consideration.
Details on the chains can be found in~\cref{sec:scanning-details}.

\begin{table}
\centering
\begin{tabular}{l | l l | l l l l l}
\hline\hline
Model & $\tilde\chi^2_\mathrm{min}$ & $\tilde \chi^2_{\rm red}$ & $\Omega_\mathrm{m,0}$ & $\OLN$ & $\log_{10}(\bar\alpha)$ & $\log_{10}(\OFP)$ & $\log_{10}(\mFP\, [\unit{eV}])$ \\
\hline
$\beta_1$ & $563.11$ & $0.973$ & $0.38^{+0.02}_{-0.02}$ & $0.79^{+0.01}_{-0.01}$ & $(-0.24)$ & $1.05^{+0.02}_{-0.02}$ & $-31.24^{+0.01}_{-0.01}$ \\
\hline
$\beta_0\beta_1$ & $562.22$ & $0.973$ & $0.28^{+0.15}_{-0.02}$ & $0.72^{+0.05}_{-0.02}$ & $-16^{+16}_{-84}$ & $-0.14^{+0.02}_{-0.02}$ & $-31.84^{+0.01}_{-0.01}$\\
$\beta_1\beta_2$ &  $562.23$ & $0.973$ & $0.28^{+0.10}_{-0.02}$ & $0.72^{+0.05}_{-0.02}$ & $-20^{+20}_{-30}$ & $39^{+61}_{-39}$ & $-12^{+30}_{-20}$  \\
$\beta_1\beta_3$ & $562.23$ & $0.973$ & $0.30^{+0.08}_{-0.04}$ & $0.70^{+0.07}_{-0.02}$ & $-1^{+1}_{-49}$ & $2^{+98}_{-2}$& $-31^{+49}_{-1}$  \\
$\beta_1\beta_4$ & $562.23$ & $0.973$ & $0.28^{+0.13}_{-0.02}$ & $0.72^{+0.03}_{-0.08}$ &  $-20^{+20}_{-30}$ & $38^{+62}_{-38}$ & $-13^{+31}_{-19}$ \\
\hline
$\beta_0\beta_1\beta_4$ & $562.19$ & $0.974$ & $0.28^{+0.15}_{-0.02}$ & $0.72^{+0.12}_{-0.05}$ & $-1_{-99}^{+3}$ & $6^{+94}_{-6}$ & $-29^{+47}_{-3}$ \\
$\beta_1\beta_2\beta_3$ &  $562.23$ & $0.974$ & $0.28^{+0.03}_{-0.02}$ & $0.72^{+0.02}_{-0.02}$  & $-65^{+64}_{-35}$ & $86^{+14}_{-87}$ & $11^{+7}_{-43}$ \\
\hline
$\Lambda$CDM & $562.25$ & $0.971$ & $0.28^{+0.03}_{-0.02}$ & $0.72^{+0.03}_{-0.02}$ & -- & -- & --\\
\hline\hline
\end{tabular}
\caption{Summary of the best fit values for the one, two and three parameter models.
To compute $\mFP$ from $\OFPN$ we use the local value of the Hubble rate,
$H_0=(73.24\pm1.74) \,\unit{km/s/Mpc} = (9.82\pm0.24)\, \unit{eV}$~\cite{Riess:2016jrr}.
The number of free parameters is given by the number of free $\beta_n$-parameters, while for
the $\Lambda\rm CDM$ model there is one free parameter.}
\label{tab:best-fit-values}
\end{table}

\begin{figure}
	\centering
	\includegraphics[scale=0.57]{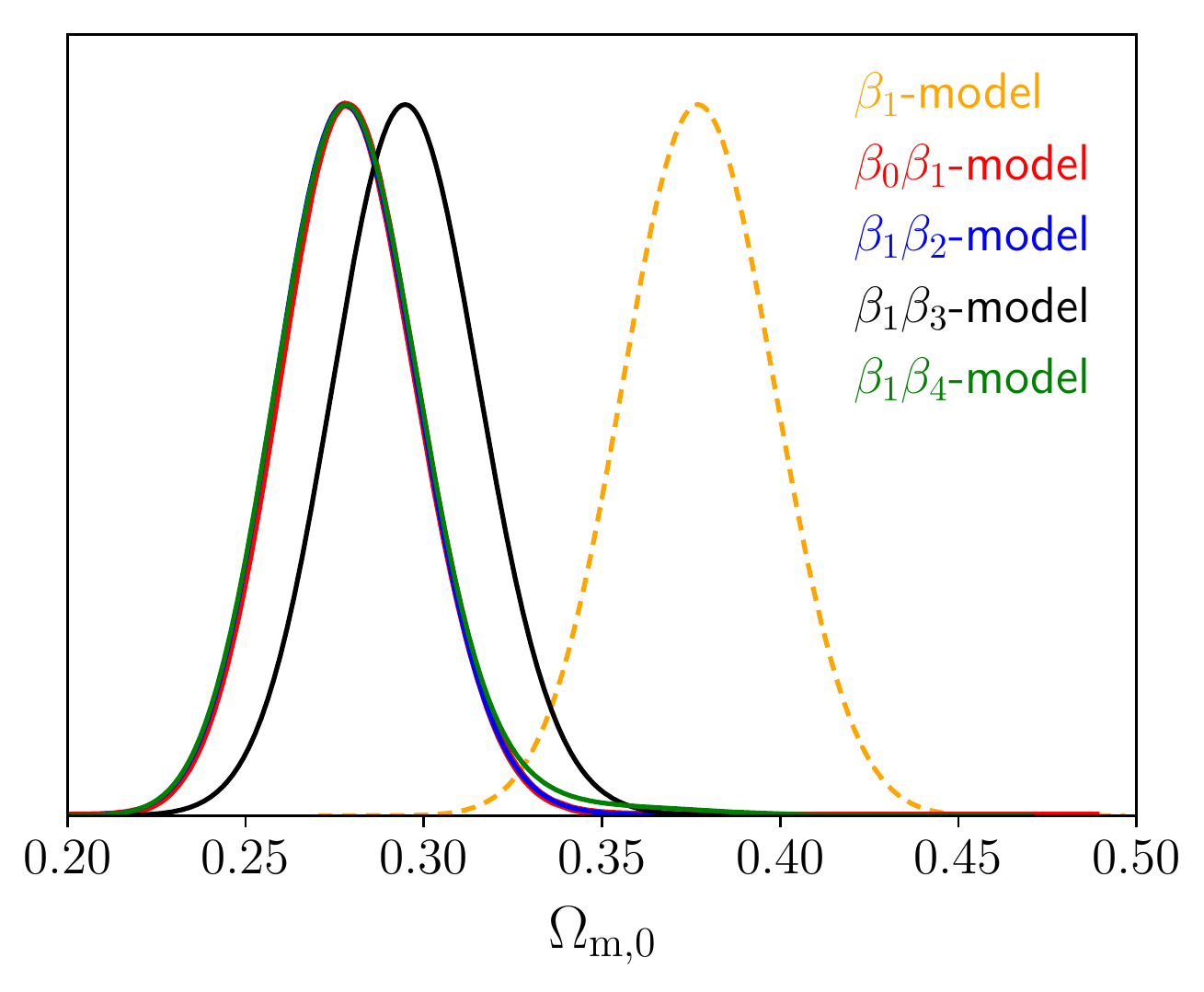}
	\includegraphics[scale=0.57]{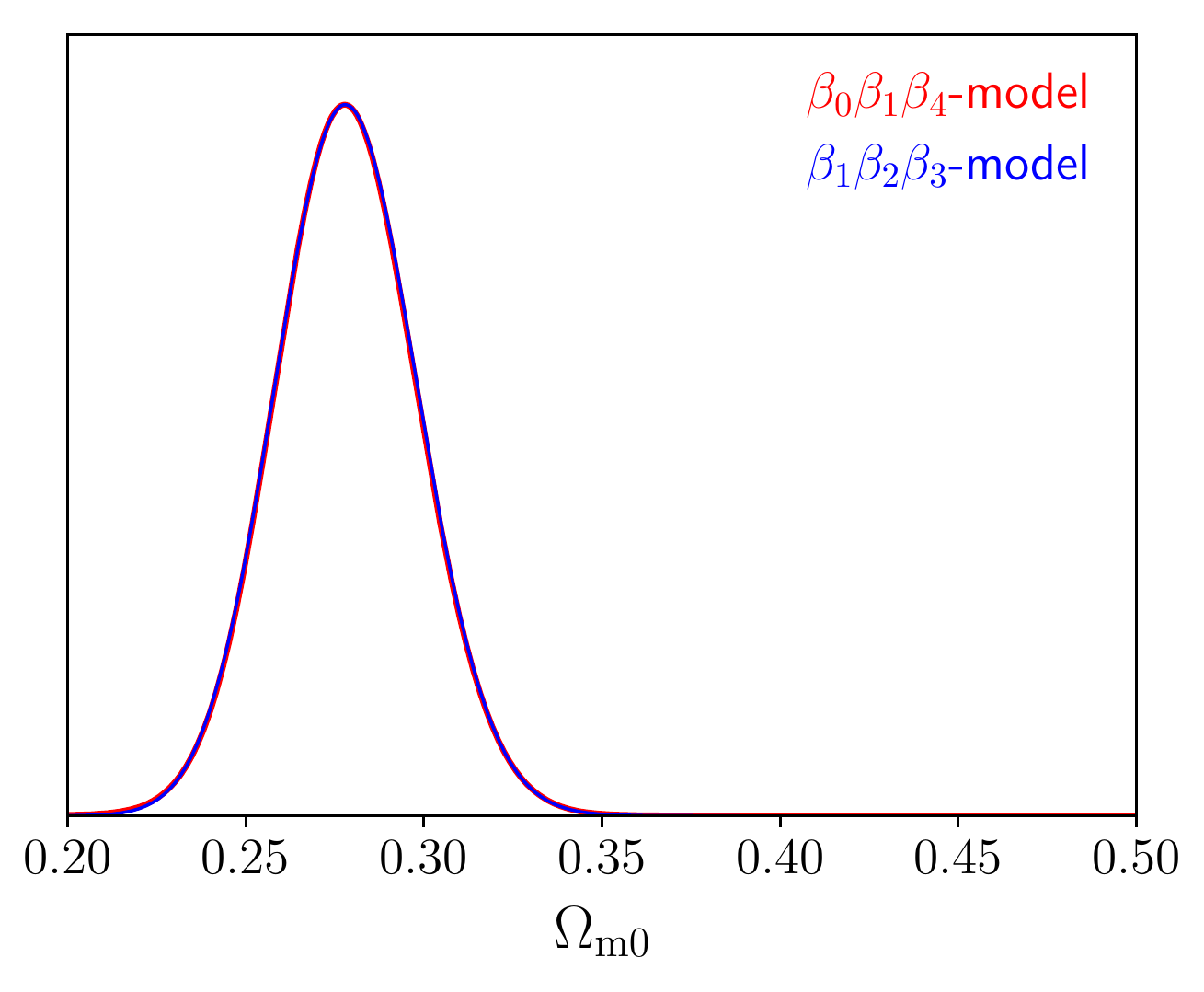}
	\caption{These figures show the marginalized posterior distribution for $\Omega_\mathrm{m,0}$. The
	left panel shows the posterior for the one- and two-parameter models, while the right
	panel shows the posterior for the three-parameter models and the full model. The
	matter energy density parameter is the exact opponent to the
	dark energy density parameter $\Omega_\mathrm{DE,0}=1-\Om$\,.}
	\label{fig:PDF-Om}
\end{figure}

Instead of discussing each model separately, let us first discuss what they have in common.
The parameters $\Om$ and $\Omega_{\Lambda,0}$ are well-constrained by
supernova data and their marginal 
posterior PDFs have a Gaussian shape. The marginal posterior PDFs for $\Om$ are depicted
in~\cref{fig:PDF-Om} for all models.
The $\beta_0\beta_1$-, $\beta_1\beta_2$-, $\beta_1\beta_4$-, and the $3$-parameter models
have roughly the same best-fit value for $\Om$
as the $\Lambda$CDM, while for the $\beta_1\beta_3$-model the value of
$\Om$ is slightly larger at the best-fit
point. The $\beta_1$-model has the largest value with $\Om=0.38\pm 0.02$.
The reason can be understood from inspecting Friedmann's equation. The
induced Dark Energy component contains a contribution that 
scales with redshift like non-relativistic matter, but with a negative sign~\cite{Babichev:2016bxi}.
This is most prominent in the $\beta_1$-model (c.f.~\cref{eq:b1-friedmann}) as it does not have
a GR-limit. Therefore, bimetric cosmology generically prefers a larger value of
$\Om$ compared to $\Lambda\rm CDM$ with beneficial impact on the
$H_0$-tension~\cite{Luben:2019yyx}.

Since
the difference of $\Om$ to unity measures the amount of Dark Energy present in the universe today,
bimetric theory needs less (or as much) Dark Energy compared to the $\Lambda$CDM model
to explain the current accelerated expansion of the universe.
Note that Dark Energy in bimetric theory, $\Omega_\mathrm{DE}=1-\Omega_\mathrm{m}$, is
not constant, but evolves in time. Most importantly, within bimetric theory the universe can be filled
with Dark Energy even in the absence of vacuum energy giving rise to self-acceleration.
The models with $\beta_0=0$ are self-accelerating and in perfect agreement with observations.
In this case, Dark Energy is composed only of interaction energy between the two metric tensors.
Only asymptotically, i.e. in the infinite future, Dark Energy approaches a constant value, 
$\Omega_\mathrm{DE,0}\rightarrow \Omega_{\Lambda,0}$.
The parameter $\Omega_{\Lambda,0}$ parametrizes the asymptotic
effective cosmological constant and is a mixture of vacuum energy and interaction energy between
the two metric tensors (unless the vacuum energy is set to zero, $\beta_0=0$).
The effective cosmological constant is well-constrained by supernova data
and the marginal posterior PDFs are Gaussian for all models. Here, we only report their best-fit value
and $1\sigma$ intervals in~\cref{tab:best-fit-values}, without explicitly showing the marginalized 
posterior PDFs.
Moreover, we can deduce that at current times Dark Energy is almost constant for all $2$-
and $3$-parameter models because
$\Omega_\mathrm{DE,0}=1-\Om \simeq \OLN$. Although we allowed for non-trivial
behavior of the bimetric models, supernova data drives all models (except the $\beta_1$-model)
into a regime where Dark Energy
behaves as a cosmological constant at current times.

Let us move to the differences between the models. The $\beta_1$-model has only one free
parameter. Data constrains the Fierz-Pauli mass to be
$\mFP=(5.71\pm 0.18)\times 10^{-32}\,\unit{eV}$. The minimal bimetric model 
gives rise to self-acceleration consistent with data when the Fierz-Pauli mass is close to the
Higuchi bound. Most notably, although the model does not possess a GR-limit ($\bar\alpha$ is
fixed by the equations of motion), it fits the data almost as good as the $\Lambda$CDM
model, with $\tilde\chi^2_\mathrm{min}\simeq 563.11$, while having the same number
of free parameters.

\begin{figure}
\centering
	\begin{subfigure}[b]{0.49\textwidth}
	\includegraphics[scale=0.7]{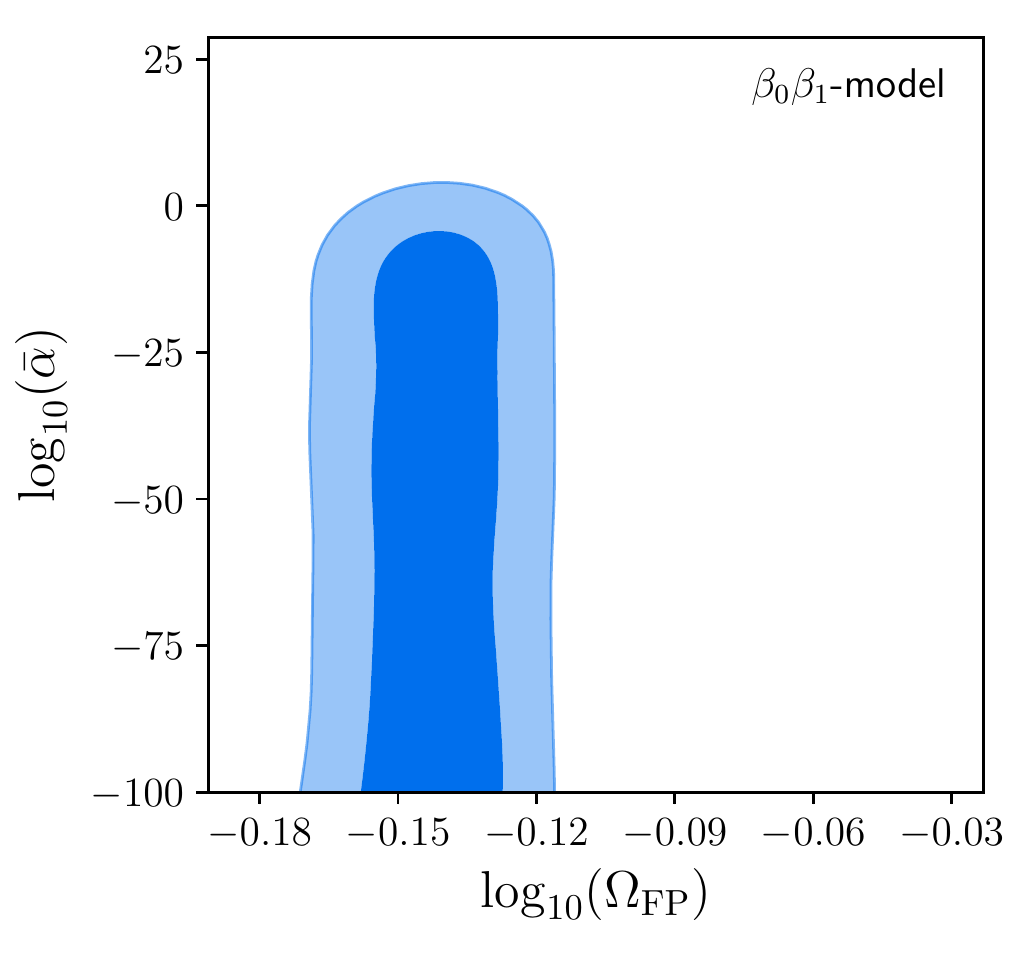}
	\caption{$68\%$ and $95\%$ c.l. for the $\beta_0\beta_1$-model}
	\label{fig:B0B1-contours}
	\end{subfigure}
	\begin{subfigure}[b]{0.49\textwidth}
	\includegraphics[scale=0.7]{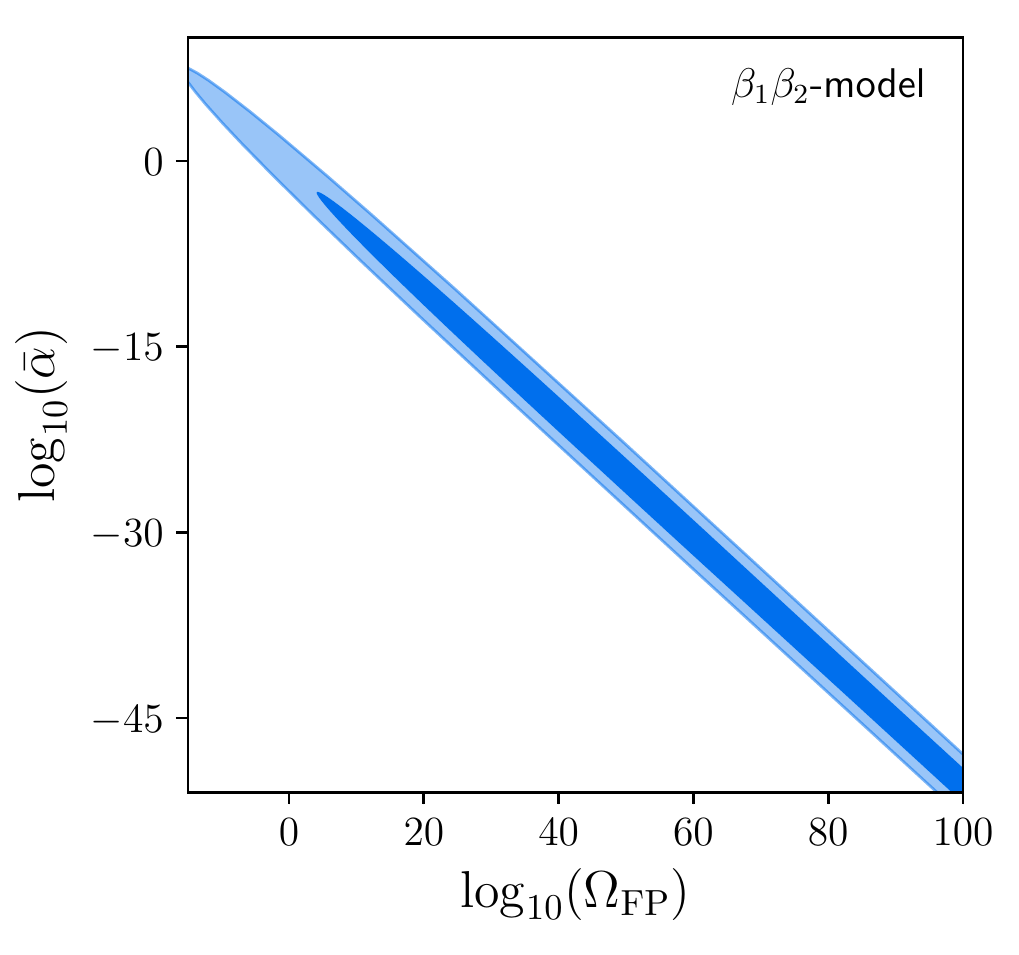}
	\caption{$68\%$ and $95\%$ c.l. for the $\beta_1\beta_2$-model}
	\label{fig:B1B2-contours}
	\end{subfigure}
	\begin{subfigure}[b]{0.49\textwidth}
	\includegraphics[scale=0.7]{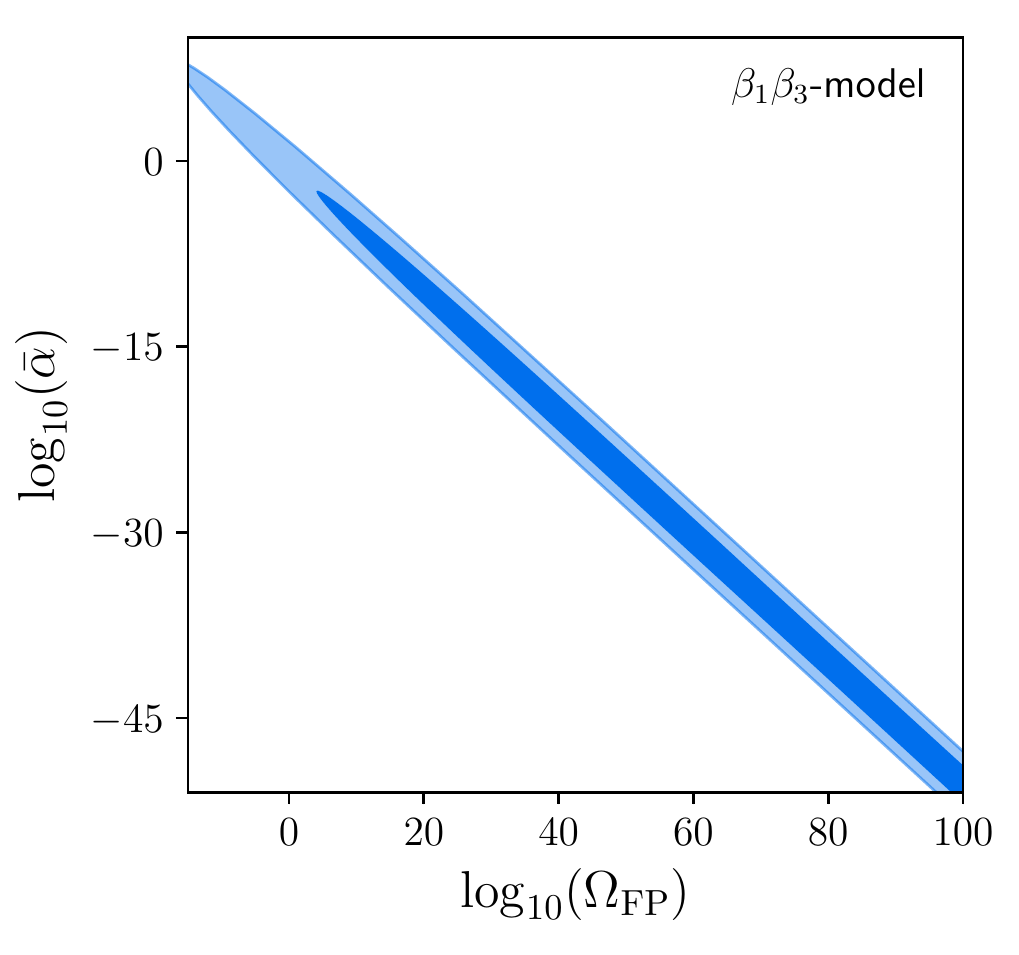}
	\caption{$68\%$ and $95\%$ c.l. for the $\beta_1\beta_3$-model}
	\label{fig:B1B3-contours}
	\end{subfigure}
	\begin{subfigure}[b]{0.49\textwidth}
	\includegraphics[scale=0.7]{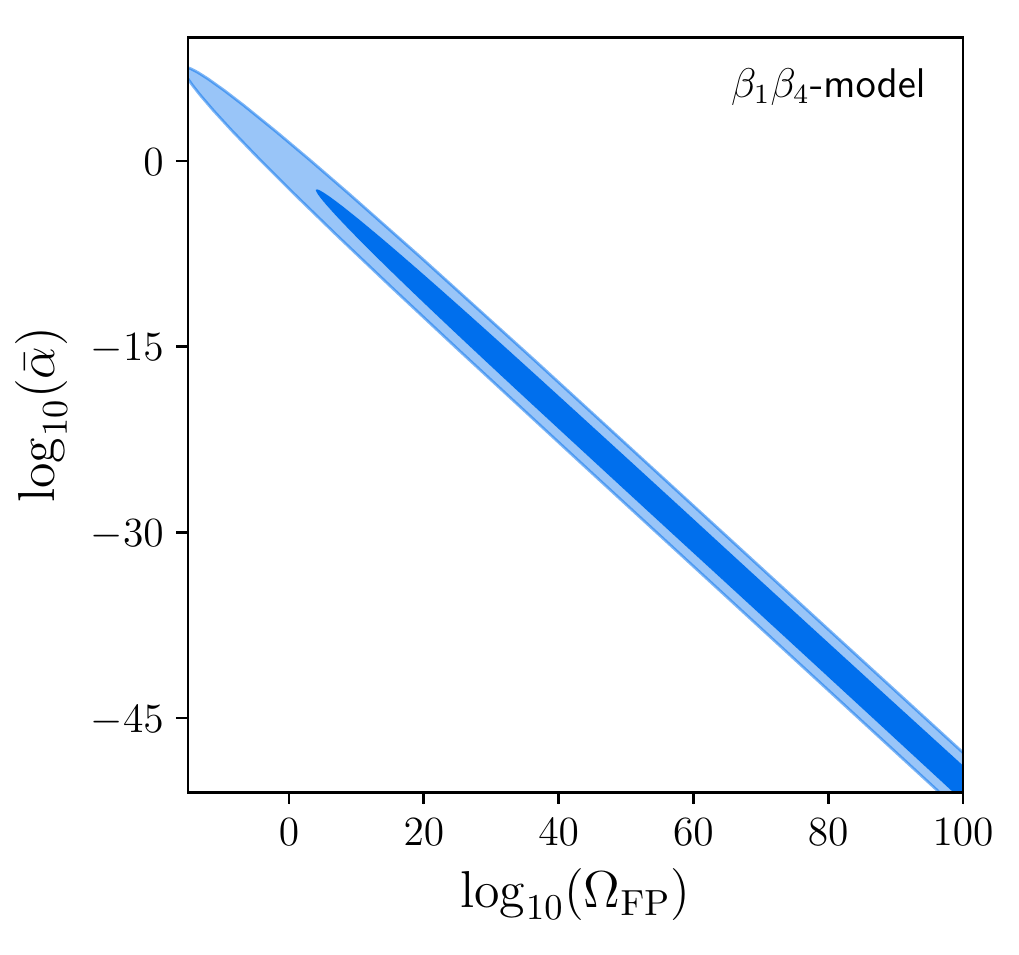}
	\caption{$68\%$ and $95\%$ c.l. for the $\beta_1\beta_4$-model}
	\label{fig:B1B4-contours}
	\end{subfigure}
	\caption{These plots show the regions of
	$68\%$ (dark blue) and $95\%$ (light blue) confidence level
	for all two parameter models
	with $\beta_1\ne 0$. They represent the two-dimensional marginal posterior PDF in the
	$\bar\alpha-\OFP$-plane.}
	\label{fig:2-param-contours}
\end{figure}

Next let us move to the two parameter models. For the $\beta_0\beta_1$-model,
the spin-$2$ mass is well-constrained to $\mFP=(1.45\pm0.05)\times 10^{-32}\,\unit{eV}$,
which lies close to the value
of the cosmological constant. Hence, this model is driven into its GR-limit,
cf.~\cref{eq:ParamReps01}.
Consequently, the mixing angle must be sufficiently small, $\bar\alpha <1$ at $1\sigma$ and
data allows the mixing angle to be arbitrarily small, cf. \ref{fig:B0B1-contours}.
The same happens in the other $2$-parameter models as can be seen
in~\cref{fig:B1B2-contours,fig:B1B3-contours,fig:B1B4-contours}.
However, the correlation between $\bar\alpha$ and $\mFP$ is different compared
to the $\beta_0\beta_1$-model. The spin-$2$ mass is not constrained by supernova data,
but only forced to not be too small ($\mFP\gtrsim 10^{-32}\,\unit{eV}$ at $1\sigma$)
for all three models.
Their correlation is determined by the previously
derived~\cref{eq:ParamReps12,eq:ParamReps13,eq:ParamReps14} and summarized
in~\cref{fig:12-param-models} with a value
for the cosmological constant as reported in~\cref{tab:best-fit-values}.
For the $2$-parameter models, our priors (specifically the upper limit on the Fierz-Pauli
mass and the lower limit on the mixing angle)
cut through a region of the physical parameter space, where the models give
a good fit to data. This, of course, is not surprising as we expect the GR-limit of these
models to give a good fit to data, which is achieved by $\bar\alpha\ll1$.
In order to decide, wether data forces these models
into their GR-limits, one needs to weaken the priors and include other observables.
This is beyond the scope of the current work.
Let us instead note that all the $2$-parameter models fit the data as good as the
$\Lambda$CDM model with $\tilde\chi_{\rm min}^2\simeq 562.2$ in all cases,
while the latter remains statistically favored due to the smaller number
of free parameters in the model. The models with
$\beta_0=0$ do not inherit vacuum energy but give rise to self-acceleration
solely due to the bimetric interaction energy.

\begin{figure}
\centering
	\begin{subfigure}[b]{0.49\textwidth}
	\includegraphics[scale=0.7]{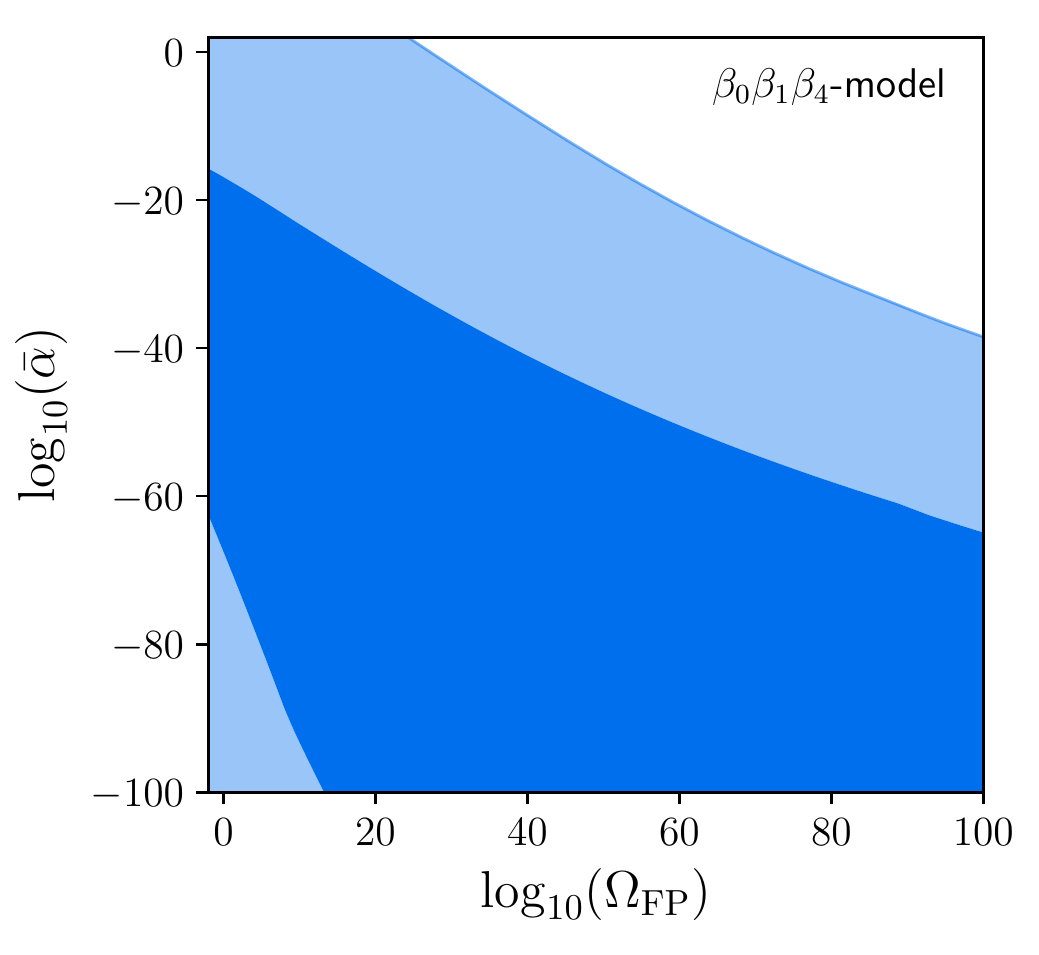}
	\caption{$68\%$ and $95\%$ c.l. for the $\beta_0\beta_1\beta_4$-model}
	\label{fig:B0B1B4-contours}
	\end{subfigure}
	\begin{subfigure}[b]{0.49\textwidth}
	\includegraphics[scale=0.7]{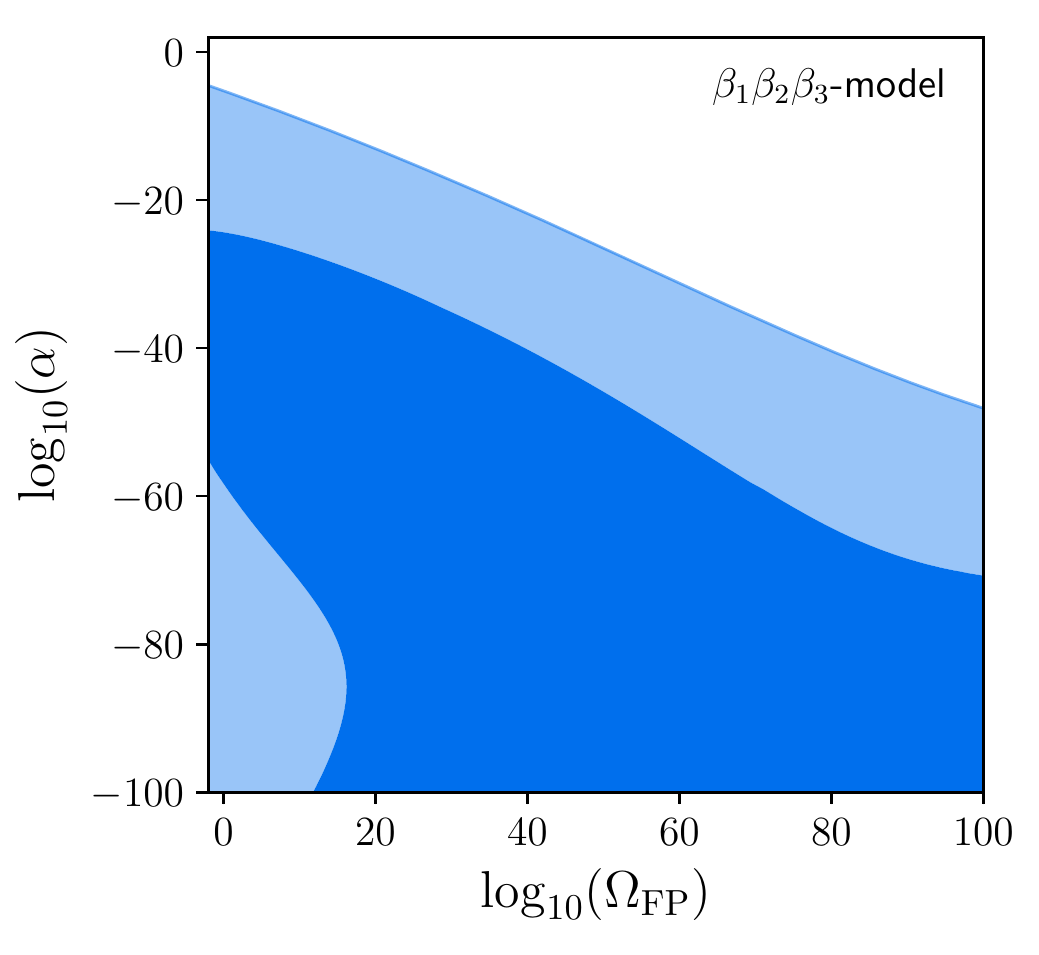}
	\caption{$68\%$ and $95\%$ c.l. for the $\beta_1\beta_2\beta_3$-model}
	\label{fig:B1B2B3-contours}
	\end{subfigure}
	\caption{These plots show the $68\%$ (dark blue) and $95\%$ (light blue) confidence levels
	for the three parameter models
	under consideration. These represent the two-dimensional marginal posterior distribution in the
	$\bar\alpha-\OFP$-plane.}
	\label{fig:3-param-contours}
\end{figure}

Let us move to the three parameter models, where the three physical parameters are
independent of each other. As for the previous models, only the parameters $\Om$ and
$\Omega_{\Lambda,0}$ are well-constrained by data, while $\bar\alpha$ and $\mFP$ remain mostly
unconstrained.
Although both three parameters are subject to completely different theoretical consistency
requirements, data selects similar parameter regions for both of them as can be seen
in~\cref{fig:3-param-contours}. For both models, the region
where $\bar\alpha$ and $\mFP$ are large are disfavored by data. For the $\beta_1\beta_2\beta_3$-
model this is obvious as this region does not satisfy our consistency requirements,
cf. \cref{fig:b1b2b3-allowed-parameter-space}.
For the $\beta_0\beta_1\beta_4$-model there is a different reason.
From~\cref{eq:ParamReps014} it follows that $\beta_0<0$ when
\begin{flalign}
	\bar\alpha^2 > \frac{\Lambda}{3\mFP^2-\Lambda}\,.
\end{flalign}
Hence, data disfavors vacuum energy to be too negative. Since vacuum energy is constant
in time ($\beta_0$ is time independent), but $y$ evolves back in time towards zero, there is a
point in the past where the Dark Energy density changes its sign and becomes negative because
$\Omega_\mathrm{DE}\rightarrow B_0$ as $y\rightarrow 0$. In principle,
$\Omega_\mathrm{DE,0}<0$
is allowed, which has to be counterbalanced by $\Om>1$.
Although theoretically consistent, this scenario is not favored by data.
The point in time, where $\Omega_\mathrm{DE}$ changes its sign, 
must be early enough for the model to be consistent with supernova data.
Since supernovae only probe times up to a redshift of $z\lesssim 1.4$, we expect cosmic 
observables that probe earlier redshifts
to put more stringent constraints on the parameter space of the 
$\beta_0\beta_1\beta_4$-model.

The region of the parameter space, where $\bar\alpha$ is small and the Fierz-Pauli mass
is close the Higuchi bound seems to be slightly disfavored by data ($2\sigma$),
cf.~\cref{fig:3-param-contours}. We believe that this is an artifact of numerical instability in that
region.
In order to see that, let us compare the value of $\tilde\chi^2$ for exemplary points in that region
with the value $\tilde\chi^2_\mathrm{min}$ for both models.
Taking $\OLN=0.72$, which is the likeliest value for both models, and exemplary
values $\bar\alpha=10^{-90}$ and $\OFP=3\cdot 10^{3}$ yields
$\tilde\chi^2=562.24$ for both models. For other exemplary parameter values we indeed
find numerical instabilities in that region as we have checked explicitly.
Hence we conclude that this region in fact is not less likely than the region enclosed
by the $1\sigma$ contour

Let us finish the discussion of three-parameter models by noting that only the upper right region
is disfavored by data, that is for large Fierz-Pauli mass $\mFP$ and a not to small
mixing angle $\bar\alpha$.
The exact GR-limit for the model is excluded
by the choice of priors (the upper limit on $\Omega_{\rm FP,0}$ and the lower limit on $\bar\alpha$),
but we expect it
to give a good fit to data as well. Wether data really forces the models into their GR-limits
needs an extended scan of the parameter space which is beyond the scope of the current work.
Besides that, a large portion of the physical parameter space is consistent with observations.

In our analysis, we use the local value of the Hubble rate,
$H_0=(73.24\pm1.74) \,\unit{km/s/Mpc}$, as reported in Ref.~\cite{Riess:2016jrr}
in order to determine $\mFP$ form $\OFP$.
However, the true value of $H_0$ is still under debate since CMB data, for instance, implies
a value of $H_0 = (67.36\pm0.54)\,\unit{km/s/Mpc}$~\cite{Aghanim:2018eyx}.
With this value we obtain a spin-2 mass of
$\mFP=(5.25\pm0.12)\, \unit{eV}$ for the $\beta_1$-model and
$\mFP=(1.33\pm0.03)\,\unit{eV}$ for the $\beta_0\beta_1$-model,
while the already large allowed mass ranges in the other models are not altered.
Hence, taking the $H_0$-tension seriously and using the global value of $H_0$
including its large errors would increase the allowed mass range.
However, since in this paper we are dealing with low-redshift data only,
we use the local value of $H_0$ instead of the global one.

\section{Conclusions and outlook}\label{sec:outlook}

We proposed a method to relate the parameters of bimetric theory to its 
physical parameters. The physical parameters are per definition entities of vacuum
solutions and since bimetric theory has several (up to four) vacua
the relation between the theory and physical parameters is a priori not unique.
However, imposing theoretical consistency requirements on the vacua and
on the expansion history of the universe, singles out a unique vacuum solution
as the true vacuum. This results in a unique relation between the theory
and physical parameters and allows to build up a dictionary between these
parametrizations.

We worked out the dictionary for all (sub-)models of bimetric theory where
the identification of the unique vacuum works analytically. These are
the models with one, two or three free interaction parameters $\beta_n$.

The physical parametrization has several advantages. 
The consistency requirements summarized in this paper for each model
and the direct physical interpretation provides intuition for the
numerical values of the parameters. This, e.g., vastly simplifies choosing
priors for data analysis.
In addition, non-linear solutions of bimetric theory that are sensitive to the
individual theory parameters
can directly be compared to linear solutions that are only sensitive to the physical
parameters.

In order to demonstrate these features of the physical parametrization
we applied our method to FLRW solutions.
As a first step towards combining various tests of gravity into a single framework
we performed a statistical analysis using supernova data to
constrain the physical parameter space of bimetric theory.

All the models that we considered show quite similar behavior with
only few exception. We find that energy density of non-relativistic matter today is
roughly the same as in GR for all models, varying from $\Omega_\mathrm{m,0}=0.28$ up
to $\Omega_\mathrm{m,0}=0.38$ for different models. The gap to the critical
energy density is filled by Dark Energy, that in bimetric theory is dynamical and either
solely due to interaction energy between the massive and the massless spin-$2$ field
(models where $\beta_0=0$) or due to a combination of vacuum and interaction energy.
Dark Energy approaches a constant value in the asymptotic future (which corresponds to the
cosmological constant of the asymptotic de Sitter spacetime) that we denote
by $\Omega_{\Lambda,0}$.
Although we allowed for nontrivial deviations from GR at late times, data
favors all the models to act like $\Lambda$CDM at late times. That is
the dynamical Dark Energy component is almost constant at late times,
$\Omega_\mathrm{DE,0}=1-\Omega_\mathrm{m,0}\approx \OLN$.
The only exception here is the $\beta_1$-model that does not have a GR-limit and is
self-accelerating.

Let us summarize the constraints on the mass, $\mFP$, and the coupling strength
to ordinary matter, $\bar\alpha$, of the massive spin-$2$ field. These constraints are
highly model dependent. For the minimal model where only $\beta_1$ is a free parameter
the coupling strength is fixed to $\bar\alpha=1/\sqrt{3}$. Data then favors a spin-$2$ mass
of $\mFP=(5.71\pm 0.18)\times 10^{-32}\,\unit{eV}$, which lies close to the Higuchi bound.
For the $\beta_0\beta_1$-model the spin-$2$ mass is constrained to
$\mFP= (1.45\pm 0.05 )\times 10^{-32}\,\unit{eV}$ while the mixing angle
is only constrained to not be too large ($\bar\alpha\lesssim1$ at $1\sigma$).
The other two parameter models show very similar behavior. The physical
parameters are degenerate
and mostly unconstrained. The coupling strength is constrained to be $\bar\alpha<1$ and the mass
$\mFP>10^{-32}\,\unit{eV}$ at $1\sigma$. For large masses, the coupling strength and the
mass are fixed by the best-fit value of the cosmological constant as
$\bar\alpha^2 \mFP^2= \kappa \Lambda$,
where $\kappa=\mathcal O(1)$ is a model dependent number\footnote{To
be precise, $\kappa=6$ for the $\beta_1\beta_2$, $\kappa=4$ for the $\beta_1\beta_3$,
and $\kappa=3$ for the $\beta_1\beta_4$-model. This follows from expanding the consistent
vacuum solutions~\cref{eq:ParamReps12,eq:ParamReps13,eq:ParamReps14} for
$\mFP^2\gg\Lambda$.}.
Note that we use the local Hubble value to determine $\mFP$,
while the allowed mass range would increase when using the global value.
Moving to the three parameter models, the strict relation between $\bar\alpha$ and $\mFP$ is
relaxed and all physical parameters are really independent. The observational constraints on
$\bar\alpha$ and $\mFP$ are weak, only the region where
$\bar\alpha^2\,\mFP^2\gg\Lambda$ is disfavored.
Bimetric theory has self-accelerating solutions even when the spin-$2$ mass is large that
are compatible with supernova data.
On the other hand, the statistical analysis performed in this paper favors the
$\Lambda\rm CDM$-model either due to the smaller number of free parameters
or due to a slightly larger likelihood.

Our results generalize the existing cosmological constraints from background observables
on bimetric theory to the entire physical parameter space.
While Refs.~\cite{vonStrauss:2011mq,Akrami:2012vf} do not distinguish between the
finite and infinite branch solution to the equations of motion,
Refs.~\cite{Akrami:2012vf,Konnig:2013gxa,Mortsell:2018mfj} compute constraints
on the interaction parameters $\beta_n$.
In Refs.~\cite{Akrami:2012vf,Konnig:2013gxa,Lindner:2020eez}
either the rescaled parametrization with $\alpha=1$ is used
or $B_n\sim\mathcal O(1)$ is assumed implying that their results
apply only to a small region of the parameter space.
As a consequence,
Refs.~\cite{vonStrauss:2011mq,Akrami:2012vf,Konnig:2013gxa,Lindner:2020eez}
test the region where the Fierz-Pauli mass is comparable to the Hubble rate today.
This parameter region is the relevant one for addressing the
$H_0$-tension~\cite{Luben:2019yyx}.
Due to the different parametrizations, it is non-trivial to relate our constraints
on the bimetric parameters to these earlier works.
Nonetheless, we can compare
the constraints on $\Om$ and find that they agree for all models that we consider.

As argued in Ref.~\cite{Aoki:2015xqa,Luben:2019yyx}, the Vainshtein
mechanism~\cite{Vainshtein:1972sx,Babichev:2013pfa} is expected to be active also in
cosmology.
It implies that deviations from GR are suppressed for energies larger than the
spin-2 mass, i.e. for $H\gg \mFP$,
while deviations are possibly testable at later times.
This however depends on $\bar\alpha$ since a small coupling to standard matter
suppresses deviations from GR at all redshifts.
That means that supernova data could in principle test this effect in the parameter region
where $\bar\alpha \gtrsim \mathcal O(1)$ and
where
$\mFP \lesssim 2\times 10^{-32}\, \rm eV$, to give a rough estimate.
This energy scale corresponds to the redshift $z\simeq 1.4$, up to which supernova data
exists.
Our analysis shows that most of this parameter region is excluded by supernova data
(to be precise, $\alpha>1$ is excluded at $95\%$ c.l.).
Hence, cosmological observables that probe higher redshifts are necessary to test the
cosmological Vainshtein mechanism.

The next step on the level of background cosmology is to include more
cosmological observables to find stronger constraints on the physical parameter space.
The constraints from background cosmology can then be combined with
constraints from other observables. While for galaxy cluster scales to galactic scales first
results were obtained~\cite{Platscher:2018voh,Enander:2015kda,Enander:2013kza,Sjors:2011iv}
they still need to be put into a single framework. Also a consistent inclusion of the
Vainshtein screening mechanism is (partly) lacking. On smaller scales, say solar system and
below, some constraints on the parameter space were derived in Ref.~\cite{Enander:2015kda}.
A comprehensive analysis including the Vainshtein screening mechanism still needs to be done.
Local and laboratory tests of gravity are reviewed in, e.g.,
Refs.~\cite{Gundlach:2005tz,Reynaud:2008yd,Will:2014kxa,Hoyle:2004cw,Adelberger:2003zx}.

The second large class of constraints comes from the perturbative level.
Although the linear perturbations on the finite branch are necessarily plagued
by a gradient
instabilities~\cite{Konnig:2014dna,Konnig:2014xva,Konnig:2015lfa,Lagos:2014lca,DeFelice:2014nja},
we can always go to a limit of bimetric theory such that the instabilities occur only
above a certain cutoff scale~\cite{Akrami:2015qga}. 
This scale is set by the Fierz-Pauli mass~\cite{Luben:2019yyx}.
In this limit bimetric theory behaves like GR and all possibly testable deviations
from the $\Lambda$CDM predictions are suppressed.
However, a gradient instability only signals a breakdown of perturbation theory.
There is good evidence that the instabilities are cured
either due to the onset of the local Vainshtein mechanism~\cite{Mortsell:2015exa} or due
to the aforementioned time-dependent analogue of the Vainshtein
mechanism~\cite{Aoki:2015xqa,Luben:2019yyx}.
Indeed, the authors of Ref.~\cite{Hogas:2019ywm} analyzed a fully non-linear but simplified
setup and did not find any instability at all.
These are promising hints that the FLRW solutions to bimetric theory are well-defined.
A new treatment of cosmological perturbations needs to be established in order to
deal with constraints coming from the perturbative level within bimetric theory.

Summarizing, in this paper we took a step towards combining various observable
constraints. There is a large portion of the physical parameter space that is consistent
with supernova data.
If the massive spin-$2$ field should also account for the observed Dark Matter abundance in
the Universe, it must be heavy (depending on the production mechanism,
$\unit{MeV}$ to $\unit{TeV}$)~\cite{Babichev:2016bxi,Chu:2017msm}.
Remarkably, our analysis shows
that a heavy spin-$2$ field is in perfect agreement with supernova data. The reason is that
the Fierz-Pauli mass and the effective cosmological constant are independent of each other
and can be of a completely different energy scale.
This opens up the possibility that bimetric theory can account for Dark Energy and simultaneously
provides Dark Matter.
While this seems to be excluded due to perturbativity bounds, answering this question requires
further study.

\section*{Acknowledgements}

M.L. acknowledges Julio A. M\' endez-Zavaleta, who joined the project at an early stage.
M.L. further thanks Georgia Pollina, Nico Hamaus and Martin Kerscher for their help
regarding the data analysis
and Angelo Caravano for useful comments on the manuscript.
The authors also thank the anonymous referee for useful comments.
This work is supported by a grant from the Max Planck Society.

\appendix

\section{Example: Tuning of the interaction parameters}
\label{sec:tuning}

In this appendix we discuss the constant and singular roots for a concrete example,
the $\beta_1\beta_2$-model.
Setting $\beta_0=\beta_3=\beta_4=0$ yields the background equation
\begin{flalign}
	3\alpha^2\beta_2 c^3+3\alpha^2\beta_1 c^2 - 3\beta_2 c - \beta_1 =0\,.
\end{flalign}
This equation represents a polynomial in $c$ of degree $3$ such that it has
up to three real-valued roots.
Instead of presenting the full solutions, let is jump to the limit $\alpha\ll1$ immediately.
We find the constant root
\begin{flalign}
	c_{\rm c} = - \frac{\beta_1}{3\beta_2} + \mathcal O(\alpha^2)\,.
\end{flalign}
In order for the root to be positive valued, we need $\beta_2<0$.
Plugging this root into the expressions for the Fierz-Pauli mass
and the cosmological constant, we arrive at
\begin{subequations}
\begin{flalign}
	\mFP^2 &= -\frac{\beta_2}{\alpha^2} + \mathcal O (1)\,,\\
	\Lambda &= -\frac{2}{3} \frac{\beta_1^2}{\beta_2} + \mathcal O(\alpha^2)\,.
\end{flalign}
\end{subequations}
For $\beta_2<0$, both quantities are positive.
This explicitly demonstrates that $\alpha\ll1$ implies $\mFP^2\gg\Lambda$ without
further tuning on a constant root.

Let us move to the singular roots, that in the limit $\alpha\ll1$ read
\begin{flalign}
	c_{\rm s\pm} = \pm\frac{1}{\alpha}-\frac{\beta_1}{3\beta_2} + \mathcal O (\alpha^2)\,,
\end{flalign}
of which $c_{\rm s+}>0$. On this singular root, the Fierz-Pauli mass and cosmological constant
are given by
\begin{subequations}
\begin{flalign}
	\mFP^2 & = \frac{4\beta_2}{\alpha^2} + \mathcal O (1)\,,\\
	\Lambda & = \frac{3\beta_2}{\alpha^2} + \mathcal O (1)\,.
\end{flalign}
\end{subequations}
Both quantities are positive valued only for $\beta_2>0$ in contrast to the constant root.
Note that $\alpha\ll1$ does not imply that the Fierz-Pauli mass is much larger than the
cosmological constant. Instead, they are of the same order of magnitude, but still 
satisfy the Higuchi bound.
In order to achieve a hierarchy between both quantities requires tuning
one of the $\beta_n$ parameters, e.g. $\beta_2=\beta_2(\alpha,\beta_1)$.

\section{Dictionary for the three parameter models}
\label{sec:submodels-dict-appenix}

In this appendix, we identify the consistent vacua
and complete the dictionary between the theory
and physical parameters for the remaining three
parameter models, but without comparing these
models to data.

\subsubsection{$\beta_0\beta_1\beta_2$ model}

For $\beta_3=\beta_4=0$, \cref{eq:cquartic} has the following roots
\begin{flalign}
	\bar\alpha_\pm = \pm \sqrt{\frac{\mFP-\Lambda+\bar\beta_2}{\Lambda-\bar\beta_2}}\,,
\end{flalign}
where $\bar\beta_2=\alpha^{-2}\beta_2$.
The root $\bar\alpha_-$ is always non-positive and we dismiss it.
The root $\bar\alpha_+$ is real-valued in the parameter range $\Lambda-\mFP^2<\bar\beta_2<\Lambda$.
Solving this relation for $\beta_2$, we can express all interaction parameters
in terms of physical parameters as
\begin{subequations}
\begin{flalign}
	\beta_0 & = \frac{\bar\alpha^2(-6\mFP^2+4\Lambda)+ (1+3\bar\alpha^4)\Lambda}{1+\bar\alpha^2}\\
	\alpha^{-1}\beta_1 & = \bar\alpha \left(\frac{3\mFP^2}{1+\bar\alpha^2} - 2\Lambda\right)\\
	\alpha^{-2}\beta_2 & = \frac{(1+\bar\alpha^2)\Lambda-\mFP^2}{1+\bar\alpha^2}
\end{flalign}
\end{subequations}
The vacuum point $\bar\alpha_+$ is well-defined in the whole parameter space, when the Higuchi bound is satisfied.
However, requiring $\beta_1>0$ imposes a bound on the parameter space,
\begin{flalign}\label{eq:b0b1b2-bound}
	3\mFP^2>2(1+\bar\alpha^2)\Lambda\,.
\end{flalign}
The left panel of \cref{fig:b0b1b34-param-space} shows the theoretically
consistent parameter space of the $\beta_0\beta_1\beta_2$-model. In the
red-shaded region~\cref{eq:b0b1b2-bound} is violated and hence unphysical.
Translating the bound on $\beta_2$ yields the same condition in the physical parameters.

\subsubsection{$\beta_0\beta_1\beta_3$ model}

\begin{figure}
	\centering
	\includegraphics[scale=0.8]{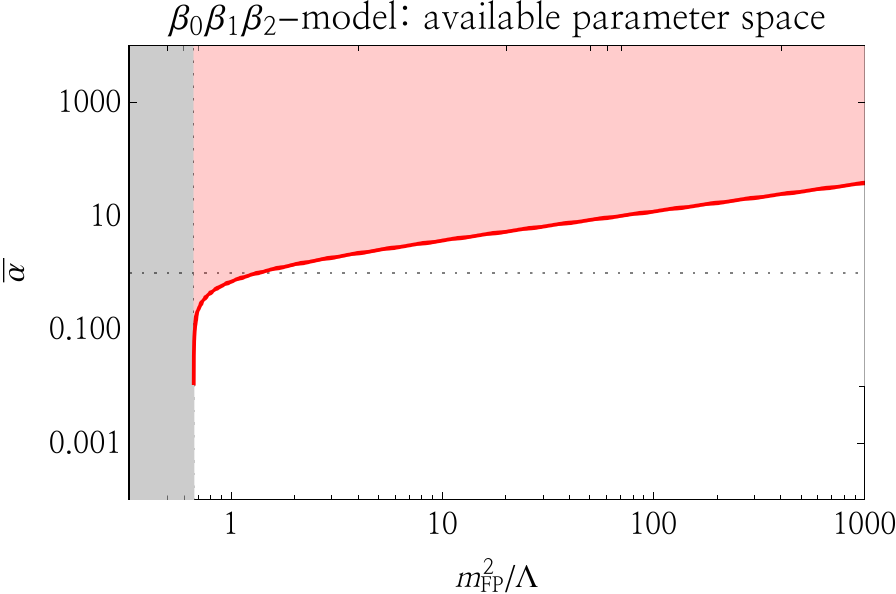}
	\includegraphics[scale=0.8]{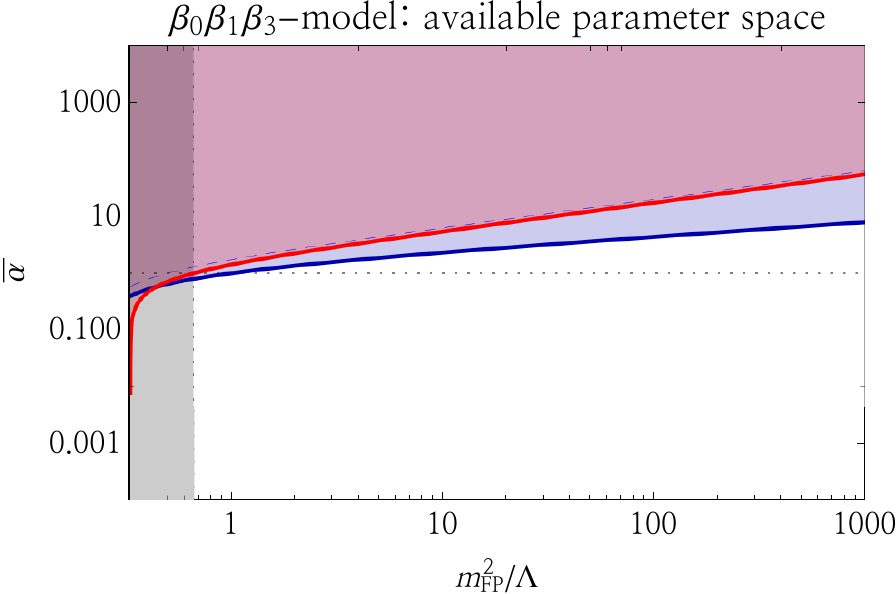}
	\caption{Left:
	The theoretically consistent parameter space for the $\beta_0\beta_1\beta_2$ model.
	In the red-shaded region the condition $\beta_1>0$ is violated.
	Right:
	The theoretically consistent parameter space of the $\beta_0\beta_1\beta_3$ model.
	The blue-shaded region is excluded because the asymptotic future is not well-defined.
	In the red-shaded region the bound $\beta_1>0$ is violated.
	The blue dashed line indicates the bound~\cref{eq:b034-consistency}.
	In both plots, the gray-shaded region indicates violation of the Higuchi bound.}
	\label{fig:b0b1b34-param-space}
\end{figure}

For $\beta_2=\beta_4=0$, the polynomial~\cref{eq:cquartic} has four roots, of which the possibly positive
ones read.
\begin{flalign}
	\bar\alpha_\pm = \sqrt{\frac{4\mFP^2-2\Lambda+\beta_0 \pm\sqrt{(4\mFP^2+\beta_0)^2-16\mFP^2\Lambda}}{2\Lambda}}\,.
\end{flalign}
The other two roots are $-\bar\alpha_\pm$ and hence non-positive in the entire 
parameter space.
Both roots are real-valued only if $\beta_0>4\mFP(\sqrt{\Lambda}-\mFP)$.
In this parameter range we find that $\bar\alpha_+$ can never be smaller than $\bar\alpha_-$.
This identifies $\bar\alpha_-$ as the unique vacuum.
It is real-valued in the parameter range $4\mFP(\sqrt{\Lambda}-\mFP)<\beta_0<\Lambda$.
Solving the expression for $\bar\alpha_-$ for $\beta_0$ yields
\begin{subequations}
\begin{flalign}
	\beta_0 &= \frac{\bar\alpha^2 (-4\mFP^2+2\Lambda)+(1+\bar\alpha^2)\Lambda}{1+\bar\alpha^2}\\
	\alpha^{-1}\beta_1 & = -\bar\alpha \frac{-3\mFP^2 + (1+\bar\alpha^2)\Lambda}{2(1+\bar\alpha^2)}\\
	\alpha^{-3}\beta_3 & = \frac{-\mFP^2+(1+\bar\alpha^2)\Lambda}{2\bar\alpha(1+\bar\alpha^2)}
\end{flalign}
\end{subequations}
where we already simplified the expressions using the bound \cref{eq:cosmobound013}.
The constraints on $\beta_0$ translate as
\begin{flalign}\label{eq:b034-consistency}
	4\mFP^2>(1+\bar\alpha^2)\Lambda\,.
\end{flalign}
Outside this parameter range, the vacuum $\bar\alpha_-$ is not well-defined.
This bound is weaker than the other bounds and represented by the blue dashed line in~\cref{fig:b0b1b34-param-space}.
The requirement $\beta_1>0$ is satisfied in the parameter region where
\begin{flalign}
	3\mFP^2>(1+\bar\alpha^2)\Lambda\,,
\end{flalign}
which is indicated by the red-shaded region in~\cref{fig:b0b1b34-param-space}.

Moving to cosmology, we expand the expression for $\rho_\mathrm{m}(y)$ around 
$y=\bar\alpha/\alpha$ and find that it only vanishes if
\begin{flalign}\label{eq:cosmobound013}
	4\mFP^2>(1+\bar\alpha^2)^2\Lambda
\end{flalign}
is satisfied. This bound is represented by the blue-shaded region.
In this parameter range, $\beta_1$ is guaranteed to be positive and represents the most stringent bound
on the parameter space.
Only if these bounds are satisfied, the $\beta_0\beta_1\beta_3$ model can give rise to a viable expansion history.
The bounds are collected in the right panel of~\cref{fig:b0b1b34-param-space}.

\subsubsection{$\beta_1\beta_2\beta_4$ model}

Setting $\beta_0=\beta_3=0$, \cref{eq:cquartic} has four roots, of which 
the two possibly positive ones are given by
\begin{flalign}
	\bar\alpha_\pm = \sqrt{\frac{3\mFP^2-\Lambda+3\bar\beta_2
	\pm \sqrt{ (3\mFP^2-\Lambda-3\bar\beta_2)^2-12\bar\beta_2\Lambda }}{6\bar\beta_2}}\,,
\end{flalign}
where $\bar\beta_2=\alpha^{-2}\beta_2$.
The other two roots are given by $-\bar\alpha_\pm$ and we neglect them.
A necessary condition for both roots to be real-valued is
\begin{flalign}\label{eq:b1b2b4-consistency1}
	3\bar\beta_2<3\mFP^2-2\sqrt{3}\mFP\sqrt{\Lambda}+\Lambda\,.
\end{flalign}
Furthermore, the root $\bar\alpha_+$ is real valued only if additionally $\beta_2>0$.
In this parameter range $\bar\alpha_-<\bar\alpha_+$ always. 
This identifies $\bar\alpha_-$ as the unique consistent vacuum of the $\beta_1\beta_2\beta_4$ model
in the viable parameter range defined by \cref{eq:b1b2b4-consistency1}.

Solving the expression for $\bar\alpha_-$ for the remaining interaction
parameter $\beta_2$ yields the following dictionary
\begin{subequations}
\begin{flalign}
	\alpha^{-1}\beta_1 &= \frac{\bar\alpha^2(-3\mFP^2+2\Lambda)+2\Lambda}{2\bar\alpha (1+\bar\alpha^2)}\,,\\
	\alpha^{-2}\beta_2 & = \frac{\bar\alpha^2 (3\mFP^2-\Lambda)-\Lambda}{3\bar\alpha^2 (1+\bar\alpha^2)}\,,\\
	\alpha^{-4}\beta_4 & = \frac{\bar\alpha^2 (-6\mFP^2+4\Lambda) +(1+3\bar\alpha^4)\Lambda}{3\bar\alpha^4 (1+\bar\alpha^2)}\,,
\end{flalign}
\end{subequations}
where we already used the bound~\eqref{eq:b124-consistency2} to simplify expressions.
The consistency requirement $\beta_1>0$ is satisfied if
\begin{flalign}
	3\bar\alpha^2\mFP^2 < 2 (1+\bar\alpha^2)\Lambda\,.
\end{flalign}
In the red-shaded region in the left panel of \cref{fig:b0b1b34-param-space} this bound is violated.

Moving to cosmology, the finite branch is well-defined if
\begin{flalign}\label{eq:b124-consistency2}
	3\bar\alpha^4 \mFP^2 < (1+\bar\alpha^2)^2 \Lambda
\end{flalign}
is satisfied.
The blue-shaded region in the left panel of~\cref{fig:b0b1b34-param-space}
indicates, where this bound is violated.

\begin{figure}
	\centering
	\includegraphics[scale=0.8]{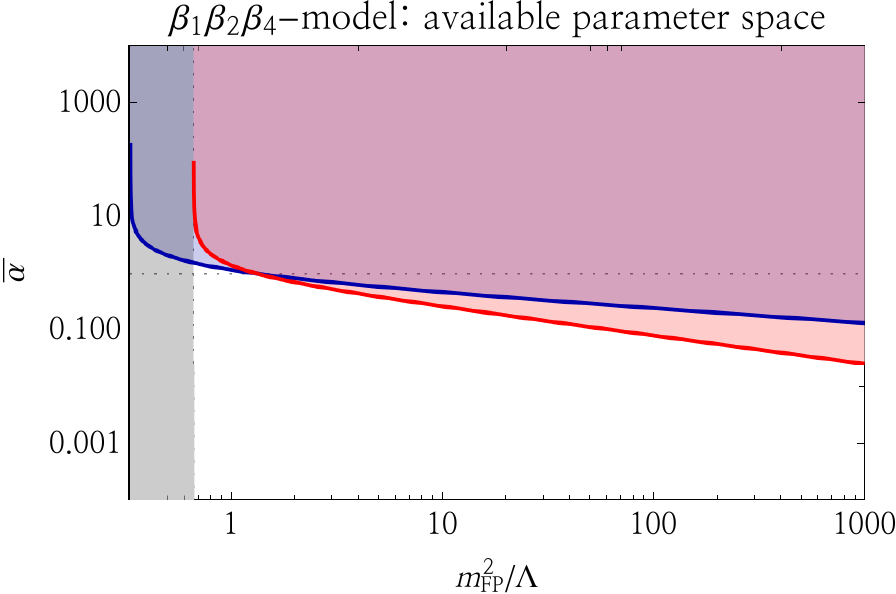}
	\includegraphics[scale=0.8]{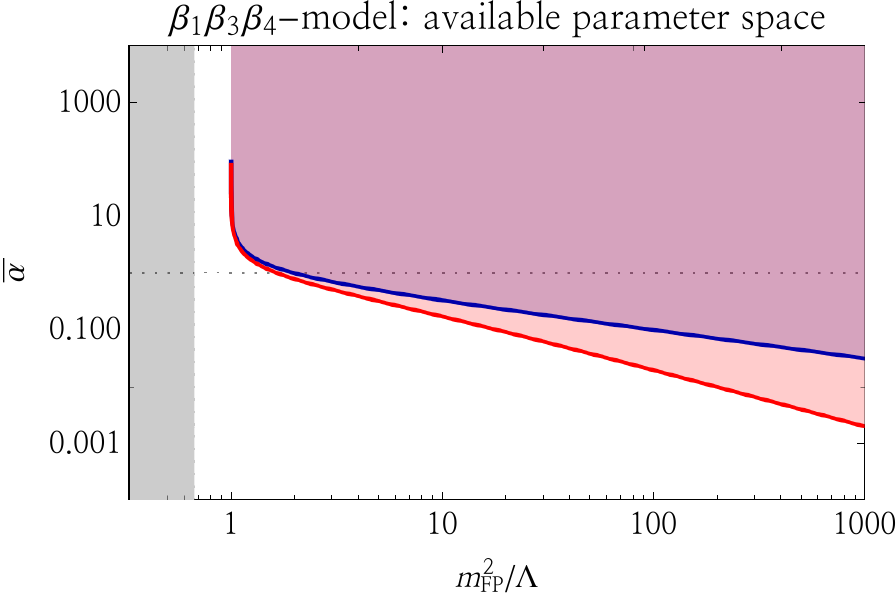}
	\caption{Left:
	The theoretically consistent parameter space for the $\beta_1\beta_2\beta_4$ model.
	In the blue-shaded region the finite branch is not well-defined, while in the red-shaded region $\beta_1>0$ is violated.
	Right: The theoretically consistent parameter space for the $\beta_1\beta_3\beta_4$ model.
	In the red-shaded region the bound $\beta_1>0$ is violated, while in the blue-shaded region the vacuum point is not well-defined.
	In both plots, the gray-shaded region indicates violation of the Higuchi bound.}
	\label{fig:b1b23b4-param-space}
\end{figure}

\subsubsection{$\beta_1\beta_3\beta_4$ model}

For $\beta_0=\beta_2=0$ the background \cref{eq:cquartic} has three roots, only one of which is possibly real-valued,
\begin{flalign}
	\bar\alpha = \frac{1}{6\bar\beta_1} \left( -\mFP^2+\Lambda + \frac{-12\bar\beta_1^2 + (\mFP^2-\Lambda)^2}{\mathcal B^{1/3}} + \mathcal B^{1/3} \right)\,,
\end{flalign}
where
\begin{flalign}
	\mathcal B = -(\mFP^2-\Lambda)^3 + 18\bar\beta_1^2 (\mFP^2+2\Lambda)
	+6\bar\beta_1 \sqrt{48\bar\beta_1^4-3(\mFP^2-\Lambda)^3\Lambda-3\bar\beta_1^2 (\mFP^4-20\mFP^2\Lambda-8\Lambda^2)}\,.
\end{flalign}
By means of analytical and numerical methods,
we find that the root is real-valued and positive if
\begin{flalign}\label{eq:b134-consistency1}
	\alpha^{-1}\beta_1 > \frac{ \sqrt{\mFP^4-8\Lambda^2 +\mFP (\mFP^2+8\Lambda)^{3/2}}}{4\sqrt{2}}\,,
\end{flalign}
which represents a nontrivial bound only if $\mFP^2>\Lambda$.
Solving the expression for $\bar\alpha$ for $\beta_1$ and plugging the result into the expressions for the other
interaction parameters yields the following map:
\begin{subequations}
\begin{flalign}
	\alpha^{-1}\beta_1 & = \frac{-\bar\alpha^2\mFP^2 + (1+\bar\alpha^2)\Lambda}{2\bar\alpha (1+\bar\alpha^2)}\,,\\
	\alpha^{-3}\beta_3 & = \frac{\bar\alpha^2 (3\mFP^2-\Lambda)-\Lambda}{2\bar\alpha^3 (1+\bar\alpha^2)} \,, \\
	\alpha^{-4}\beta_4 & = \frac{-4\bar\alpha^2\mFP^2+(1+\bar\alpha^2)^2\Lambda}{\bar\alpha^4 (1+\bar\alpha^2)}\,,
\end{flalign}
\end{subequations}
The consistency bound~\eqref{eq:b134-consistency1} on the parameters translates into
\begin{flalign}
	4(1+\bar\alpha^2)^3 \Lambda^2> \bar\alpha^2 ( 2(1+\bar\alpha^2)\Lambda + \mFP^2 )^2\,.
\end{flalign}
The blue-shaded region in the right panel of \cref{fig:b1b23b4-param-space} indicates, where this bound is violated.
The bound $\beta_1>0$ is satisfied in the parameter region where
\begin{flalign}
	\bar\alpha^2 \mFP^2 < (1+\bar\alpha^2)\Lambda\,.
\end{flalign}
The red-shaded region in the right panel of \cref{fig:b0b1b34-param-space} represents
the region of the parameter space, where $\beta_1>0$ is violated.
Note that for this submodel, the small strip between the Higuchi bound and $\mFP^2=\Lambda$
is not excluded by out consistency requirements.

\section{Details of the scan}\label{sec:scanning-details}

\begin{table}
\centering
\begin{tabular}{l | l | l }
\hline\hline
Model & Scanning parameters & $\mathcal R-1$\\
\hline
$\beta_1$ & $\Om$ & $0.003$ \\
\hline
$\beta_0\beta_1$ & $\log_{10}(\bar\alpha)$, $\log_{10}(\OFP)$  & $0.001$ \\
$\beta_1\beta_2$ & $\log_{10}(\OFP)$, $\OL$ & $0.008$\\
$\beta_1\beta_3$ & $\log_{10}(\OFP)$, $\OL$ & $0.004$ \\
$\beta_1\beta_4$ & $\log_{10}(\OFP)$, $\OL$ & $0.002$ \\
\hline
$\beta_0\beta_1\beta_4$ & $\log_{10}(\bar\alpha)$, $\log_{10}(\OFP)$, $\OL$ & $0.01$ \\
$\beta_1\beta_2\beta_3$ & $\log_{10}(\bar\alpha)$, $\log_{10}(\OFP)$, $\OL$ & $0.008$ \\
\hline\hline
\end{tabular}
\caption{For different models we used a different set of free scanning parameters 
over which the MCMC runs. For the full model we choose to use $\sin^{-1}(B_{1,4})$
because it allows to scan many orders of magnitude for both positive and negative values.
In addition, we report the value of the Gelman-Rubin factor $\mathcal R$.
For sufficient convergence, the factor should be $\mathcal R-1\lesssim 0.01$.}
\label{tab:scanning-params}
\end{table}

In order to be explicit, in this appendix we report how we set up our MCMCs
and comment on their convergence.
For each model, we set up three independent MCMCs with different starting points.
The free scanning parameters vary from model to model and are summarized
in~\cref{tab:scanning-params}.

For the $\beta_1$-model, the MCMCs run over $\Omega_\mathrm{m,0}$ with a variance
of $0.01$ and the three starting point $\{0.1,0.5,0.8\}$. We stop the chain after $15000$ steps and
remove $100$ steps as burn-in. This yields a Gelman-Rubin factor of
$\mathcal R-1\simeq 0.003$. Since we have only one parameter this
already provides enough statistics and the chains have nicely converged.

For the $\beta_0\beta_1$-model, the MCMCs run over $\log\bar\alpha$ with starting points
$\{1,-10,-50\}$ and over $\log\Omega_\mathrm{FP}$ with starting points
$\{1,1,1\}$. Stopping the chain
after $150000$ steps and removing $6500$ steps as burn-in yields a Gelman-Rubin
factor of $\mathcal R-1\simeq 0.001$.

For the remaining two parameter models, we use the same scanning parameters and starting
points. $\log_{10}\OFP$ starts at $\{1,50,80\}$ and $\OL$ at $\{0.5,0.1,0.8\}$.
In each case we remove $500$ steps as burn-in.
The Gelman-Rubin factors for the three Markov chain for each model is
$\mathcal R-1\approx 0.008$ for $\beta_1\beta_2$, $\mathcal R-1\simeq0.004$ for $\beta_1\beta_3$,
and $\mathcal R-1\approx0.002$ for $\beta_1\beta_4$ signaling sufficient convergence.

Moving two the three parameter models, we only considered the two extreme cases and
used the same scanning parameters. The parameter $\log_{10}\bar\alpha$ starts
at $\{-1,-70,-70\}$, the parameter $\log_{10}\OFP$ at $\{1,10,70\}$, and $\OL$ starts at
$\{0.7,0.7,0.7\}$ for both models. We remove $500$ steps as burn-in. 
The Gelman-Rubin factor is $\mathcal R-1\approx 0.01$ for the $\beta_0\beta_1\beta_4$-model and
$\mathcal R-1\approx 0.008$ for the $\beta_1\beta_2\beta_3$-model.

Summarizing, all our Markov chains suggest sufficient convergence and provide
enough statistics for parameter inference.

\bibliographystyle{unsrt}
\bibliography{BimetricPhysicalParameterSpace}

\end{document}